\documentclass[twocolumn]{aastex63}
\usepackage{amsmath}

\received{June 1, 2019}
\revised{January 10, 2019}
\accepted{\today}
\submitjournal{ApJ}

\shorttitle{[$\mathrm{O~\textsc{iii}}$] emitters in D4}
\shortauthors{Shi et al.}

\begin{document}

\title{Narrowband Imaging of a $z=3.24$ Protocluster: Insights from [O III] Emitting Galaxies}

\correspondingauthor{Ke Shi}
\email{shike.astroph@gmail.com}

\author{Ke Shi}

\affiliation{School of Physical Science and Technology, Southwest University, Chongqing 400715, China}

\author{Jun Toshikawa}

\affiliation{Nishi-Harima Astronomical Observatory, Center for Astronomy, University of Hyogo, 407-2 Nishigaichi, Sayo-cho, Sayo, Hyogo 679-5313, Japan}

\author{Xianzhong Zheng}
\affiliation{Tsung-Dao Lee Institute and Key Laboratory for Particle Physics, Astrophysics and Cosmology, Ministry of Education, Shanghai Jiao Tong University,
Shanghai, 201210, China}

\author{Zheng Cai}
\affiliation{Department of Astronomy, Tsinghua University, Beijing 100084, China}

\author{Dongdong Shi}
\affiliation{Center for Fundamental Physics, School of Mechanics \& Optoelectronic Physics, Anhui University of Science \& Technology, Huainan 232001, China}



\begin{abstract}
We present a narrowband imaging on a spectroscopically confirmed protocluster ``D4UD01'' at $z=3.24$ using CFHT/WIRCam. We identify a sample of 24 [$\mathrm{O~\textsc{iii}}$] emission line galaxies in the field, which forms a large overdensity in the protocluster region. The protocluster is expected to evolve into a Virgo-like cluster by $z=0$. Utilizing multiwavelength data, we derive the physical properties of these [$\mathrm{O~\textsc{iii}}$] emitters and find they are medium mass normal star-forming galaxies ($\sim10^{10}$M$_\sun$) roughly following the star-forming main sequence. The [$\mathrm{O~\textsc{iii}}$] emitters trace an overdensity spatially offset from that of photometric-redshift and quiescent galaxies, suggesting these distinct galaxy populations may inhabit dark matter halos that formed at different epochs. A comparative analysis of [$\mathrm{O~\textsc{iii}}$] emitter properties shows similar characteristics in both protocluster and field environments. This protocluster likely represents an evolved structure that has progressed beyond its peak star-formation phase, although our limited sample size may prevent detection of subtle environmental effects.

\end{abstract}



\section{Introduction} \label{sec:intro}
Decades of research have shown that environmental conditions play a fundamental role in shaping galaxy formation and evolution. The local universe shows a clear environmental dichotomy in galaxy morphology: galaxy clusters are characterized by early-type galaxies such as ellipticals, while low-density fields are populated primarily by late-type galaxies like spirals and irregulars. This ``morphology-density'' relation has been observed by numerous studies up to $z\sim1$ \citep[e.g.,][]{Dressler80,Dressler97, Goto03, Kauffmann04, Mei23, Shi24a}. Moreover, analysis of stellar populations indicates that cluster galaxies experienced short but intense periods of star formation, terminated by rapid quenching \citep[e.g.,][]{Stanford98,Thomas05,Snyder12,Mart18}. Therefore, galaxy clusters serve as valuable natural laboratories for investigating how environment shapes galaxy formation and evolution.

To fully comprehend the formation mechanisms and early evolutionary states of galaxy clusters, it is essential to study protoclusters -- the progenitors of present-day galaxy clusters. Protoclusters, forming at redshifts $z\gtrsim2$, are not virialized structures characterized by their high-density environments and diverse populations of galaxies. Many protoclusters exhibit enhanced star-formation activities relative to the general field \citep[e.g.,][]{Steidel05,Koyama13,Hayashi16,Wang16,Shimakawa18,Ito20,Shi20,Polletta21,Perez24,Staab24,Toshikawa24b}, probably due to more frequent galaxy interactions and mergers, larger gas reservoirs, or higher gas accretion rates within their deep gravitational potentials. While these findings support the proposed reversal of ``star formation-density'' relation at high redshift in dense environments \citep{Lemaux22,Shi24a,Taamoli24}, there are also many protoclusters that exhibit similar properties as the field \citep[e.g.,][]{Cucciati14,Lemaux14,Lemaux18,Shi191,Koyama20,Polletta21,Perez23}. The absence of environmental effects may reflect the protocluster's evolutionary phase: either before environmental influences become dominant or after the peak star-formation epoch of the protocluster. Although, the small sample sizes in some of these studies could also account for the observed patterns. What is more, observations have revealed substantial populations of quiescent galaxies in many protoclusters \citep[e.g.,][]{Steidel05,Kubo13,Lemaux14,Lemaux18,Zavala19,Ando20,Shi21, McConachie22,Ito23, Shi24b, Tanaka24, Jin24,Naufal24}, challenging our previous understanding of early galaxy evolution in dense environments. 

In recent years, advances in multi-wavelength observation -- spanning optical, near-infrared, submillimeter, and radio wavelengths -- have significantly improved our ability to detect and characterize protoclusters. However, a systematic search and extensive study of protoclusters still faces several challenges. As protoclusters span tens of megaparsecs \citep{Chiang13} and often exhibit complex, multi-component structures, their identification requires precise redshift measurements and wide-field surveys. Some studies used distant radio galaxies or quasars as signposts of overdense regions to identify protoclusters, which are confirmed by follow-up spectroscopy \citep[e.g.,][]{Pentericci00,Kurk04,Kashikawa07,Venemans07,Hatch11,Kuiper11,Hayashi12,Wylezalek13,
Cooke14,Adams15}. However, these are probably biased tracers of the underlying matter distribution since many studies found no association between these signposts and protoclusters \citep[e.g.,][]{Husband13,Uchiyama18,Shi192}. ``Blank-field'' surveys from extensive spectroscopy have also been conducted to discover dozens of  protoclusters \citep[e.g.,][]{Steidel05,Toshikawa12,Cucciati14,Cucciati18,Lee14,Lemaux14,Lemaux18,Dey16,Toshikawa16,
Jiang18,Toshikawa20,Hung25}, many of which are found by preselecting overdense regions traced by star-forming galaxies such as Lyman-alpha Emitters (LAEs) or Lyman-Break Galaxies (LBGs). While the number of protoclusters are growing rapidly, our understanding of these systems remains limited in two critical areas: 1) how these massive structures evolved over cosmic time to become the galaxy clusters we observe today, and 2) how the dense protocluster environment influenced the formation, growth, and properties of their member galaxies \citep{Overzier16,Alberts22}. 

In this paper, we present a follow-up study of a protocluster in the Deep 4 (D4) field of the Canada-France-Hawaii-Telescope Legacy Survey (CFHTLS). The protocluster, D4UD01, was first discovered by a systematic search for protoclusters based on LBG overdensities at $z\sim3-5$ by \cite{Toshikawa16}. A significant surface density of LBGs was found in the D4 field. Follow-up spectroscopy confirmed at least five galaxies at $z=3.24$ within 2 physical Mpc with one another, suggesting it will become a virialized cluster at $z=0$ by comparing with simulations \citep{Toshikawa16}. To investigate in detail the physical properties of protocluster galaxies and their environmental effects, \cite{Shi21} conducted a multiwavelength study of massive galaxies in this protocluster using photometric redshift (photo-$z$). They also identified a large photo-$z$ galaxy overdensity ($3.0<z<3.4$) in the field which aligns with the LBG overdensity. Notably, massive quiescent galaxies are found to be largely concentrated in the protocluster, suggesting an accelerated mass assembly and early onset of environmental quenching. However, the photo-$z$ galaxies selected by \cite{Shi21} are massive ones ($\sim10^{11} M_\odot$) that are highly incomplete at low-mass end. Thus the sample could potentially be biased against the genuine structure of the protocluster. 

To further probe the faint-end slope of the stellar mass function of star forming galaxies, we carry out a deep narrowband imaging to search for [$\mathrm{O~\textsc{iii}}$] emitting galaxies in the protocluster field. Compared to commonly used emission line galaxies such as LAEs, [$\mathrm{O~\textsc{iii}}$] emitters represent a more evolved population with higher stellar masses and are less sensitive to dust attenuation \citep{Finkelstein09,Suzuki16}.  While LAE selection can be biased against dusty systems due to resonant scattering of Ly$\alpha$ photons, [$\mathrm{O~\textsc{iii}}$] emission more effectively traces the general star-forming galaxy population at high redshifts. Many observational studies have leveraged [$\mathrm{O~\textsc{iii}}$] emission lines to identify and characterize protoclusters at high redshifts \citep[e.g.,][]{Maschietto08,Forrest17,Wen22,Daikuhara24}. At the redshift of D4UD01, [$\mathrm{O~\textsc{iii}}$] emissions fall within the Near-IR narrowband filter of CHFT/Wide-field InfraRed Camera (WIRCam), enabling an independent study of how less massive star-forming galaxies coexist with massive quiescent galaxies in the protocluster environment.

This paper is organized as follows. In Section \ref{sec2}, we present the narrow-band imaging of the D4UD01 field and our method of photometry. In Section \ref{sec3} we describe our methods of selecting the [$\mathrm{O~\textsc{iii}}$] emission line galaxies and measurement of their physical properties. In Section \ref{sec4} we investigate the spatial distribution of these galaxies and give an estimate of the protocluster mass. In Section \ref{sec5} we discuss the environmental effects on galaxy properties and explore the angular separation between different galaxy populations. A summary of our results is given in Section \ref{sum}. Throughout this paper, we use the Planck cosmology from \cite{Planck20}. All magnitudes are given in the AB system \cite{Oke83}. Distance scales are given in comoving units unless noted otherwise.

\section{Data and Photometry} \label{sec2}
In this section, we describe the data used in this work, as well as our methods of generating a multi-band photometric catalog. 
\subsection{Data} \label{data}
In the fall of 2021 and 2022, we conducted narrowband imaging surrounding the protocluster candidate D4UD01 in the D4 field, one of the four CFHTLS deep fields (CFHT Program: 21BS06; P.I.: Ke Shi). The pointing center is [$\alpha$, $\delta$]$=$[22:14:13, -18:01:58]. The narrowband imaging was taken with WIRCam on the CFHT using the $H_2S1$ filter ($\lambda_c=2.122\,\micron$, $\Delta \lambda=0.032\,\micron$). The $H_2S1$ filter samples redshifted [$\mathrm{O~\textsc{iii}}$]$\lambda 5007$ emission in the range $z=3.24\pm0.03$, spanning a line-of-sight distance of 54.4 Mpc.

The individual exposure time is 180\,s with dithering to fill in CCD chip gaps. We discard frames with seeing greater than $1.\arcsec2$. The total exposure time of the remaining frames is 6.2 hours. The mosaic image has a native pixel scale of 0.$\arcsec$3.

The WIRCam data are reduced using the pipeline \textit{SIMPLE} (Simple Imaging and Mosaicking PipeLinE) \citep{Wang10}, an IDL based data reduction package. The basic procedure includes flat fielding, background subtraction, cosmic rays removal, and  removal of instrumental features. The pre-processed frames are further astrometrically calibrated against Sloan Digital Sky Survey (SDSS)
Data Release 17 (DR17) catalog \citep{Abdurro22}. The distortion corrected and projected images are then combined in $SIMPLE$ to create deep mosaic images. The photometric calibration is performed against the CFHTLS-T0007 $K_S$ band catalog \citep{Bielby12} by choosing a sample of point sources with $16<K_S<19$. We confirm the median value of magnitude difference $H_2S1-K_S$ is zero.

We trim the final mosaic image removing the area near the edges with less than 30\% of the maximum exposure time, which results in a final effective area of 378 arcmin$^2$. The average seeing of the final mosaic is $0.\arcsec9$ with a 5$\sigma$ limiting magnitude of 22.5 AB. 

In addition to the new $H_2S1$ data, we use the publicly available multiwavelength data including the deep optical $ugriz$ images from the CFHTLS Deep Servey \citep{Gwyn12} and the near-IR $JHK_S$ bands from WIRCam Deep Survey (WIRDS) \citep{Bielby12}. We also use the Spitzer IRAC 3.6\,$\micron$ and 4.5\,$\micron$ data from the NMBS-II IRAC survey \citep{Annunziatella18}. The pixel scale of CFHTLS and WIRDS images is 0.$\arcsec$186 while for IRAC data it is 0.$\arcsec$558. In order to perform photometry consistently on the multiband images later, we resample the $H_2S1$ image to have the same pixel scale of CFHTLS and WIRDS images. The broad band images are then trimmed to have the identical dimension to the $H_2S1$ image. The photometric depth and image quality are summarized in Table \ref{table1}. It is noted that the IRAC data is fairly shallow (maximum exposure time of only $\sim$1 hour for each channel) with large photometric uncertainties, and thus they have relatively weak constraints in our analysis.

\begin{deluxetable*}{cccc}
\tablecaption{Data Set \label{table1}}
\tablehead{
\colhead{Band} & \colhead{Instrument} & \colhead{Limiting magnitude \tablenotemark{$\dagger$}} & \colhead{FWHM} \\
\colhead{} & \colhead{} & \colhead{(5$\sigma$,AB)} & \colhead{($\arcsec$)}
}

\startdata
$u$ & MegaCam/CFHT & 27.02 & 0.90\\
$g$ & MegaCam/CFHT & 27.48 & 0.80\\
$r$ & MegaCam/CFHT & 27.11 & 0.70\\
$i$ & MegaCam/CFHT & 26.72 & 0.70\\
$z$ & MegaCam/CFHT & 25.84 & 0.70\\
$J$ & WIRCam/CFHT & 24.83 & 0.60\\
$H$ & WIRCam/CFHT & 24.33 & 0.60\\
$K_S$ & WIRCam/CFHT & 24.29 & 0.60\\
$H_2S1$ & WIRCam/CFHT & 22.50 & 0.90 \\
3.6 $\micron$ & IRAC/Spitzer & 22.27 & 1.86 \\
4.5 $\micron$ & IRAC/Spitzer & 22.29 & 1.75 \\
\enddata
\tablenotetext{\dagger}{5$\sigma$ limiting magnitude measured in a 2$\arcsec$ diameter aperture for the CFHT data, while for the Spitzer data it is measured in a 3$\arcsec$ aperture.}
\end{deluxetable*}

\subsection{Photometry} \label{photometry}
We create a multiwavelength photometric catalog as follows. First, we smooth the WIRDS $JHK_S$ images and CFHTLS $ugriz$ images to match the largest point-spread function (PSF) of the $H_2S1$ band data. To do so, the radial profile of the PSF in each band is approximated by a Moffat function with the corresponding FWHM. A noiseless convolution kernel between the low and high resolution images is then derived using the Richardson-Lucy deconvolution algorithm \citep{Richardson:72}. Finally, each image is convolved with its respective kernel to match the PSF of the $H_2S1$ image. 

We perform source detection and photometric measurements on the multiband images using the SExtractor software \citep{Bertin96} in duel mode. The $H_2S1$ image is used for detection while flux measurements are done on other PSF-matched CFHTLS and WIRDS bands. The fluxes and magnitudes are measured in a 2$\arcsec$ diameter aperture for these bands. For sources not detected in certain bands, we use the 2$\sigma$ limiting magnitude to give the upper limits.

As for the IRAC images, the broad PSFs lead to severe blending for photometric measurements which then requires source deblending. In order to obtain accurate and unbiased measurement of fluxes and colors on the IRAC images, we utilize the T-PHOT software \citep{Merlin15,Merlin16}. T-PHOT performs ``template-fitting'' photometry on the low-resolution image using the information of high-resolution image and catalog. In our case, the $H_2S1$ band image and catalog are used as the input priors of T-PHOT while the low-resolution IRAC images are analyzed to obtain corresponding photometry.

Finally, all the catalogs are combined together to make a multiwavelength catalog. In this work, we only consider sources with $H_2S1$ magnitudes smaller than 22.5 (i.e., $>$5$\sigma$ detection limit). In the end, 7,676 sources are selected in the final catalog.

\section{Sample Selection} \label{sec3}

\subsection{Selection of Emission Line Candidates}
\begin{figure}[ht!]
\epsscale{1.2}
\plotone{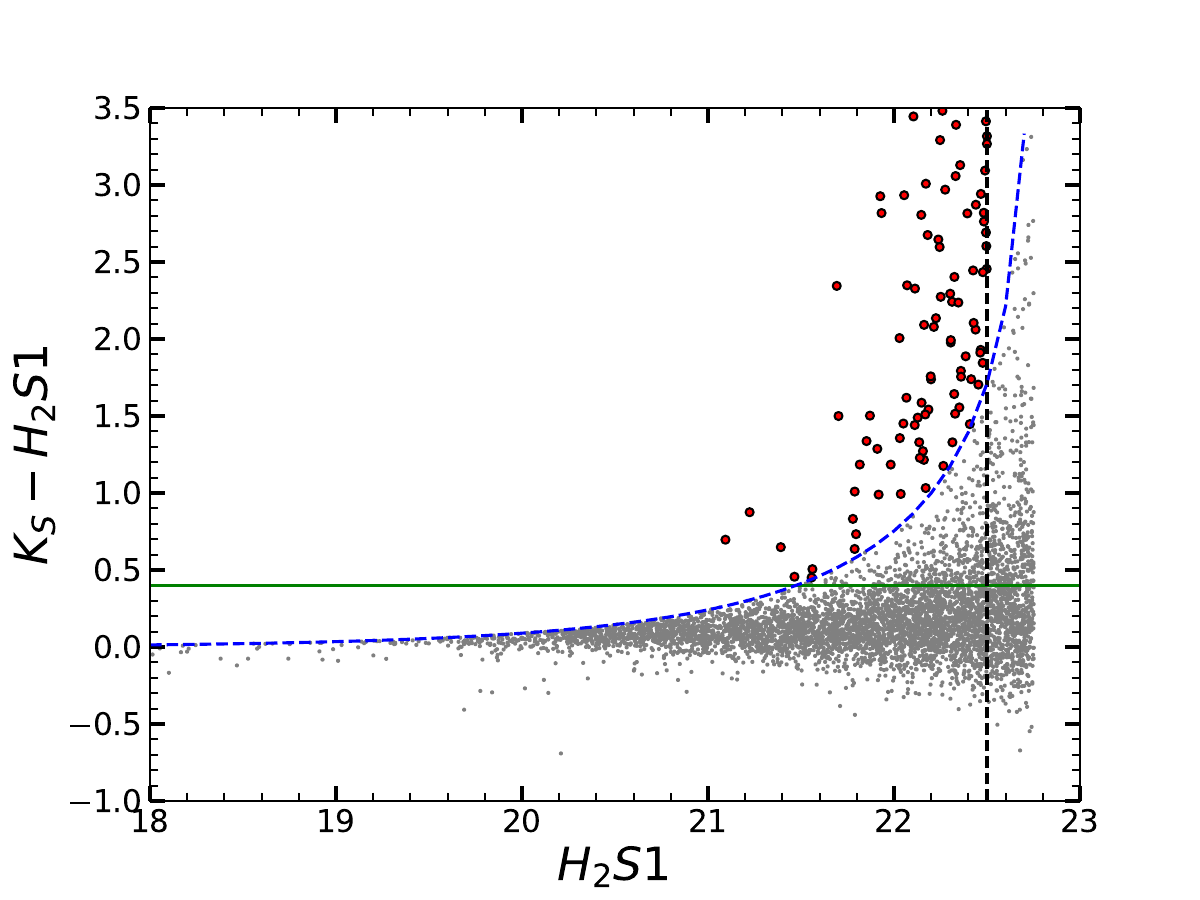}
\caption{
  Color-magnitude diagram for selecting emission line galaxies. Gray points denote all the sources detected in the narrow band image while red circles show the emission line candidates which satisfy our selection criteria. The green line represents the narrow band excess cut of EW $=30$ \AA. The blue dashed line corresponds to the selection criterion defined by Equation (1). The vertical dashed line denotes the 5$\sigma$ limiting magnitude in the narrow band image.
}
\label{figure1}
\end{figure}

Emission line galaxies are selected by strong color excess of the narrow-band with respect to the  broad-band. In this work, we use $K_S-H_2S1$ color selection following the narrow band excess defined by \citet{Bunker95}: 

\begin{equation} 
K_S-H_2S1>-2.5 \textnormal{log} \left[ 1-\frac{\Sigma\sqrt{\sigma^2_{K_S}+\sigma^2_{H_2S1}}}{f_{H_2S1}}\right],
\end{equation} 
where $\Sigma$ denotes the significance of the narrow band excess relative to a 1$\sigma$ photometric error. $\sigma^2_{K_S}$ and $\sigma^2_{H_2S1}$ are the sky noises of $K_S$ and $H_2S1$ band images, and $f_{H_2S1}$ is the narrow band flux density. We adopt $\Sigma=3$ to select emission line candidates. In addition, we also require $K_S-H_2S1>0.4$ which corresponds to a rest-frame equivalent width (EW) cut of EW$>$30 \AA. Objects with SExtractor parameter ``CLASS$\_$STAR''$>0.9$ are also removed to reduce the contamination of stars. Based on these criteria, in total 130 objects brighter than 5$\sigma$ limiting magnitudes are selected as emission line candidates. Figure \ref{figure1} shows our selection of these candidates.

\subsection{Photometric Redshift and Spectral Energy Distribution fitting} \label{sedfit}
The $H_2S1$ filter not only selects [$\mathrm{O~\textsc{iii}}$] emitting galaxies at $z=3.24$, but can also capture H$\alpha$ ($\lambda 6563$ \AA) emitters at $z=2.24$. To further determine the redshift and physical properties of the selected emission line candidates, we perform spectral energy distribution (SED) fitting with the broad-band photometry ($ugrizJHK_S[3.6][4.5]$) using the CIGALE software \citep{Noll09,Boquien19}. CIGALE is a versatile SED-fitting software that can build composite stellar population models from various single stellar population models, star formation histories, dust attenuation laws, etc. It is based on an energy balance principle assuming the energy emitted by dust in the infrared is equal to the energy absorbed by dust in UV-optical. The model templates are then fitted to the observed SEDs of galaxies from far-UV to the near infrared, and photometric redshift as well as physical properties are estimated using a Bayesian analysis. 

As for the SED templates, we use the stellar population synthesis models of \cite{BC03}, \cite{Calzetti00} reddening law where the E(B-V) values range from 0.1 to 2.0 in steps of 0.1. We also assume \cite{Chabrier03} initial mass function and metallicity values are allowed to vary between sub-solar ($Z= 0.004$) and solar ($Z= 0.02$). The star formation history is modelled with a delayed $\tau$ functional form  of SFR $\propto t\times \textrm{exp}[-t/\tau]$ where the star-forming time scale ranges from 0.1 to 10\,Gyr. Nebular emission is also included with ionization parameter log$U$ ranging from $-$4.0 to $-$1.0 in step of 0.5. The AGN template from \cite{Dale14} is also used to account for the possible dust emissions, where the AGN fraction is set to be $0\sim1$ in steps of 0.1. The input redshifts are set to be between 0.1 and 6.0 in steps of 0.1. 

To further remove possible contaminants and obtain a reliable photo-$z$ sample, we visually inspect these sources and remove those with potential contamination in the photometry, including those severely blended with nearby bright sources and near the borders of the images. To perform a more reliable SED-fitting, we also discard sources that are detected in less than three bands. In the end, 75 objects are selected as our final sample of emission line galaxies. 

\begin{figure}[ht!]
\plotone{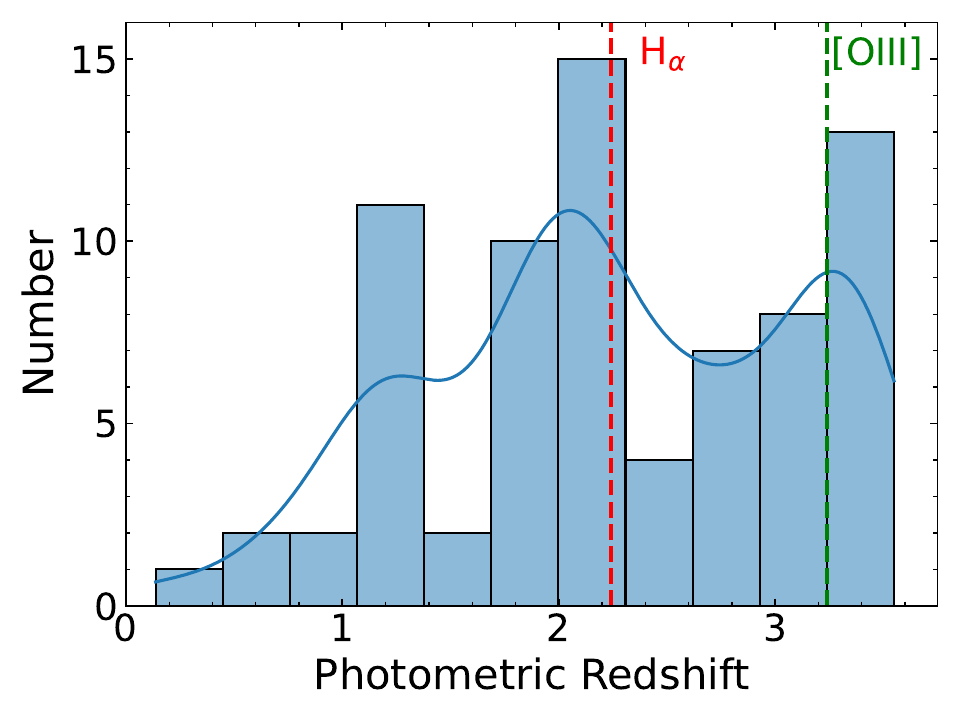}
\caption{
  Photometric redshift distribution of the 75 emission line galaxies. The smooth curve shows the underlying probability density distribution. The vertical dashed lines mark the locations of H$\alpha$ emissions (red) at $z=2.24$ and [$\mathrm{O~\textsc{iii}}$] emissions (green) at $z=3.24$, respectively.
}
\label{figure2}
\end{figure}

Figure \ref{figure2} shows the photo-$z$ distribution of the 75 emission line galaxies. We can clearly see two prominant peaks located near the corresponding redshifts of H$\alpha$ emission ($z=2.24$) and [$\mathrm{O~\textsc{iii}}$] emission ($z=3.24$), indicating that these two types of emission line galaxies dominate the galaxy populations in our sample. There is also another peak at $z\sim1.3$, which  likely corresponds to [$\mathrm{S~\textsc{iii}}$] emission at $\lambda=9069$ \AA \citep{An13}. 

\begin{figure*}[ht!]
\plottwo{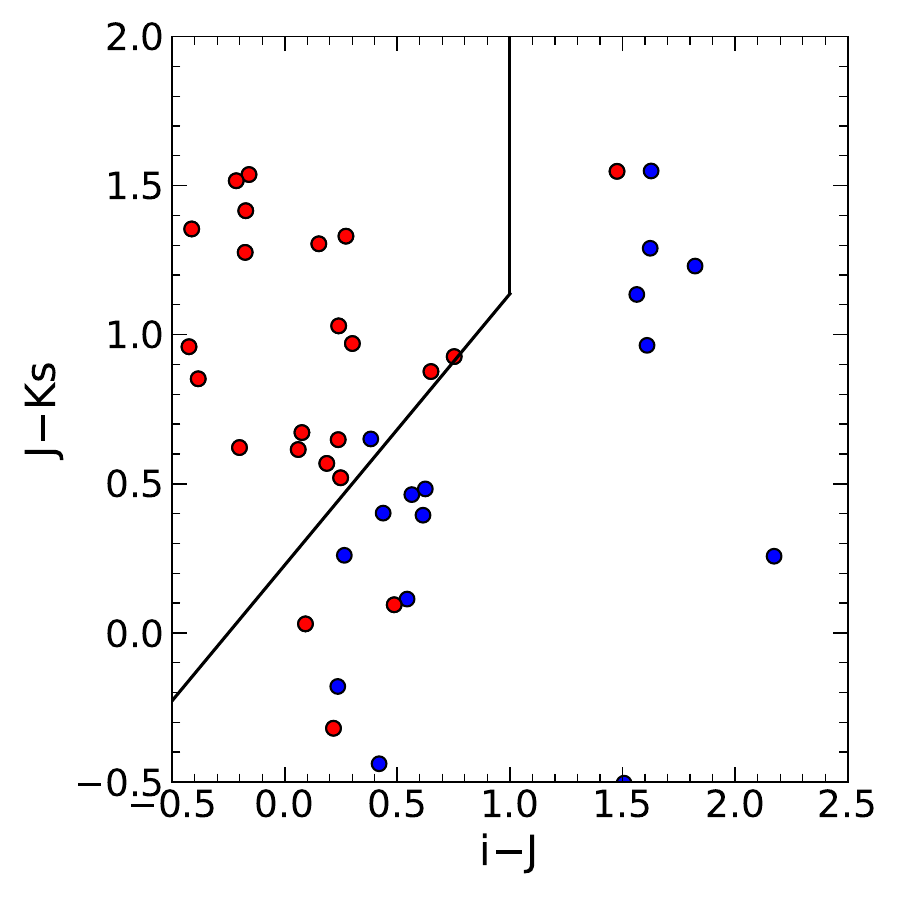}{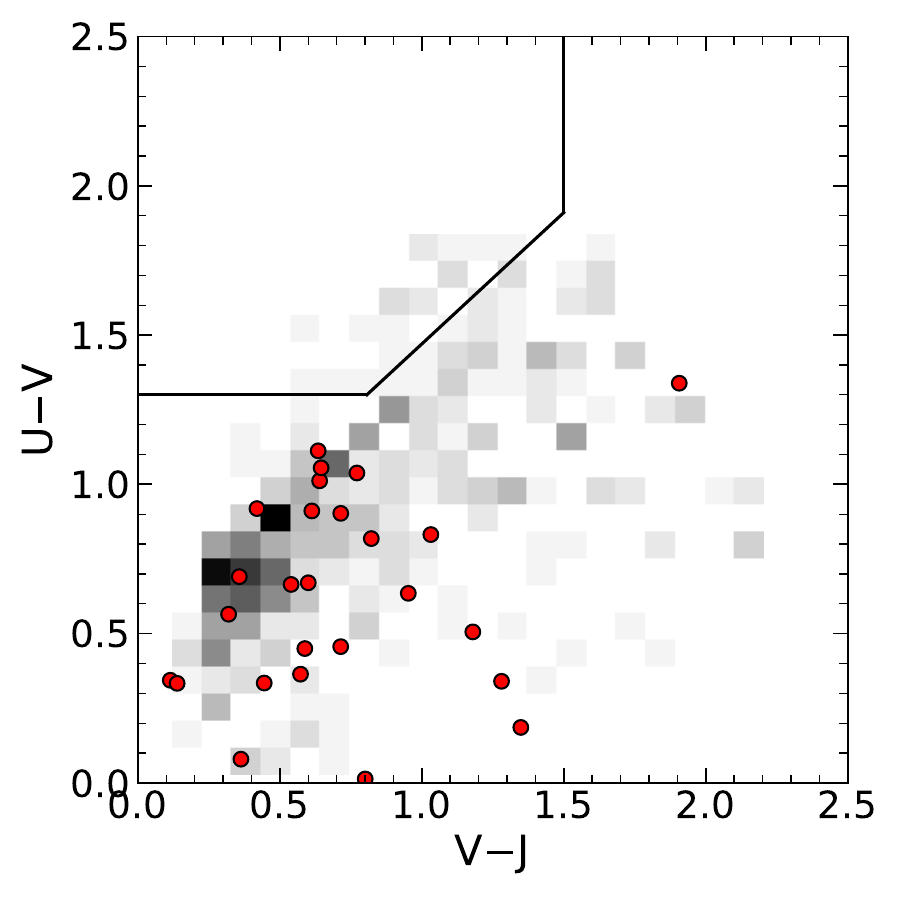}
\caption{Left: $i-J$ vs. $J-K_S$ color-color diagram. The red circles are the photo-$z$ selected [$\mathrm{O~\textsc{iii}}$] emitters while the blue circles are other emission line galaxies at lower redshifts. The solid lines represent the criteria selecting [$\mathrm{O~\textsc{iii}}$] emitters at $z>3$ \citep{Tadaki13,Suzuki15}. Right: Rest-frame $UVJ$ color-color diagram of the selected [$\mathrm{O~\textsc{iii}}$] emitters. The solid lines are the cut used to define quiescent galaxies in \cite{Muzzin13}. The gray shades show the distribution of photo-$z$ galaxies at $3.0<z<3.4$ in \cite{Shi21}. 
}
\label{figure3}
\end{figure*}

The average error of the photometric redshifts is $\Delta z_{phot}\sim 0.4$, we therefore take sources with $2.8<z_{phot}<3.6$ as potential [$\mathrm{O~\textsc{iii}}$] emitting galaxies at $z=3.25$. This leaves us with 27 candidates to further study in the following sections. We note that five of these [$\mathrm{O~\textsc{iii}}$] emitters were also found in the photo-$z$ sample ( $3.0<z_{phot}<3.4$) analyzed by \cite{Shi21}. The other [$\mathrm{O~\textsc{iii}}$] emitters are not in the photo-$z$ sample, which is mainly due to the fact that they are selected in a broader redshift range and are fitted using different SED settings. To further verify our photo-$z$ selection, in the left panel of Figure \ref{figure3} we show the $i-J$ versus $J-K_S$ diagram for separating [$\mathrm{O~\textsc{iii}}$] emitters at $z>3$ and H$\alpha$ emitters at $z<3$ as used in \cite{Tadaki13,Suzuki15}. The $iJK_S$ color criteria are devised to capture the Balmer/4000\AA ~break of galaxies at each redshift, and are also verified by spectroscopic observations \citep{Tadaki13}.
We plot only the sources with reliable color measurements. It can be seen that the majority of our photo-$z$ selected [$\mathrm{O~\textsc{iii}}$] emitters satisfy the color selection criteria, which further justify our photo-$z$ measurements. 

In the right panel of Figure \ref{figure3}, we examine the galaxy population of our [$\mathrm{O~\textsc{iii}}$] emitters in the rest-frame $U-V$ versus $V-J$ color-color diagram ($UVJ$ diagram). The $UVJ$ diagram is widely used to separate quiescent galaxies from star-forming galaxies, where the galaxies are bimodally distributed and the two populations can be separated by a color cut \citep[e.g.,][]{Labbe05,Williams09,Brammer11,Muzzin13}. For comparison, we also plot the distribution of the photo-$z$ galaxies at $3.0<z<3.4$ in the same field from \cite{Shi21}. It is clear that all of our [$\mathrm{O~\textsc{iii}}$] emitters are located in the star-forming regime, which further confirm the star-forming nature of these [$\mathrm{O~\textsc{iii}}$] emitters.

To estimate the physical properties of the selected 27 [$\mathrm{O~\textsc{iii}}$] emitters more accurately, we fix their redshifts at $z=3.24$ and refit their SEDs with CIGALE using finer parameter spaces. The CIGALE derived physical properties such as  stellar mass, star-formation rate, dust attenuation E(B-V), UV slope $\beta$ are listed in Table \ref{table2}. The typical error of stellar mass is $\sim$0.2 dex while for SFR is $\sim$0.3 dex.\\

\noindent{\it Sample Contamination ~~} 
We consider the possible contaminants in our sample here. First, it is known that AGN can contribute to the observed [$\mathrm{O~\textsc{iii}}$] emissions, especially for bright sources. Unfortunately, since there is no available X-ray or radio data in the D4 field, it is difficult to determine the AGN contribution to the [$\mathrm{O~\textsc{iii}}$] lines we detected. Therefore we resort to the SED-fitting results to examine the possible AGN components in our sample. Using \cite{Dale14} AGN model, the AGN fraction (defined as the fraction of IR luminosity from AGN to
total IR luminosity) of the 27 galaxies are estimated, among which we find three have AGN fraction $>$0.7 while others all have values smaller than 0.2. Therefore we  exclude the three sources in our following analysis and take the remaining 24 galaxies as our final sample. Further discussion about possible AGN contribution in our sample can be found in Section \ref{EWL}.

Our [$\mathrm{O~\textsc{iii}}$] emitters could also be contaminated by H$\beta$ line at $\lambda=4861$\AA. However, the bandwidth of the $H_2S1$ filter is too narrow to cover both H$\beta$ and [$\mathrm{O~\textsc{iii}}$] simultaneously at the protocluster redshift. In addition, \cite{Khostovan15} studied the redshift evolution of [$\mathrm{O~\textsc{iii}}$]+H$\beta$ luminosity function and found that at $z=2.23$, [$\mathrm{O~\textsc{iii}}$] emissions dominate the luminosity function beyond L$_\mathrm{[O~\textsc{iii}]}>10^{42}$ erg s$^{-1}$. \cite{Suzuki16} also estimated the fraction of H$\beta$ emitters in their [$\mathrm{O~\textsc{iii}}$] sample at $z=2.23$, finding a very small contamination rate of 3\%.  Since  all of our sample galaxies have L$_\mathrm{[O~\textsc{iii}]}>10^{42}$ erg s$^{-1}$ (see Section \ref{EWL}), and that the [$\mathrm{O~\textsc{iii}}$] emission line becomes more prominant at higher redshifts \citep[e.g.,][]{Faisst16,Backhaus24}, we argue that the fraction of H$\beta$ emitters in our sample is negligible. 

\subsection{\textrm{$[\mathrm{O~\textsc{iii}}]$} Luminosity and Equivalent Width} \label{EWL}
The narrow-band survey allows us to measure the line flux of the emission line galaxies and thus estimate the luminosity of the [$\mathrm{O~\textsc{iii}}$] emission (L$_\mathrm{[O~\textsc{iii}]}$) as well as the equivalent width (EW). These quantities can be obtained from the narrow-band (NB) $H_2S1$ and broad-band (BB) $K_S$ flux densities, which are defined as:
\begin{equation}
f_{\textrm{NB}}=f_\textrm{c}+F_{\textrm{line}} / \vartriangle_{\textrm{NB}},
\end{equation}
\begin{equation}
f_{\textrm{BB}}=f_\textrm{c}+F_{\textrm{line}} / \vartriangle_{\textrm{BB}},
\end{equation}
where $f_\textrm{c}$ is the continuum flux density and $F_{\textrm{line}}$ is the line flux. $\vartriangle_{\textrm{NB}}$ and $\vartriangle_{\textrm{BB}}$ are the the FWHMs of the NB and BB filters, respectively. Solving the above equations, the continuum flux density, line flux and rest-frame equivalent width can be expressed as:
\begin{equation}
f_{\textrm{c}}=\frac{f_\textrm{BB}-f_\textrm{NB}(\vartriangle_{\textrm{NB}}/\vartriangle_{\textrm{BB}})}{1-\vartriangle_{\textrm{NB}}/\vartriangle_{\textrm{BB}}},
\end{equation}
\begin{equation}
F_\textrm{line}=\vartriangle_{\textrm{NB}}\frac{f_\textrm{NB}-f_\textrm{BB}}{1-\vartriangle_{\textrm{NB}}/\vartriangle_{\textrm{BB}}},
\end{equation}
\begin{equation}
\textrm{EW}_\textrm{rest}=\frac{F_\textrm{line}}{f_\textrm{c}(1+z)}.
\end{equation}
The [$\mathrm{O~\textsc{iii}}$] luminosity is then calculated as L$_\mathrm{[O~\textsc{iii}]}=4\pi D_L^2 F_\textrm{line}$, where $D_L$ is the luminosity distance at $z=3.24$. In addition to [$\mathrm{O~\textsc{iii}}$] luminosity, we also derive UV luminosity at rest-frame 1600\AA (L$_{1600}$) from the best-fitting results of CIGALE. The above physical properties are listed in Table \ref{table2}.

Our sample exhibits strong emission lines. The rest-frame equivalent widths of the [$\mathrm{O~\textsc{iii}}$] emitters have a median value of 342\AA, after excluding three sources that were not detected in the continuum image. These strong emission line galaxies are frequently found at $z>3$ with lower metallicity and higher ionizing parameters, which are likely undergoing intense star formation that lie well above the star-forming main sequence \citep[e.g.,][]{Forrest17,Onodera20,Tran20,Tang21}. However, the strong emission line galaxies reported in the literature usually have low stellar mass ($\log(M_\ast/$M$_\sun) \sim8-9$), in comparison with our sample that has a median stellar mass of $\log(M_\ast/$M$_\sun) \sim10$. As we will see in Section \ref{env}, our [$\mathrm{O~\textsc{iii}}$] emitters roughly follow the main sequence, indicating they are normal star-forming galaxies. The large EWs and high stellar masses of [$\mathrm{O~\textsc{iii}}$] emitters have also been found by \cite{Wen22} at the same redshift. While AGN activity could contribute to the high EWs, we cannot accurately determine the AGN contribution without X-ray or spectroscopic data. We note \cite{Wen22} cross-matched their sample with Chandra X-ray observations and found only one AGN among their 34 [$\mathrm{O~\textsc{iii}}$] sources. That AGN has an [$\mathrm{O~\textsc{iii}}$] luminosity of 10$^{44}$ erg s$^{-1}$, significantly higher than our sample where the majority sources have [$\mathrm{O~\textsc{iii}}$] luminosities below 10$^{43}$ erg s$^{-1}$.  It has also been suggested that AGNs mainly contribute to the brightest end of the [$\mathrm{O~\textsc{iii}}$] luminosity function, especially at $>$10$^{43}$ erg s$^{-1}$ \citep{Khostovan15}, therefore we conclude that our sample predominantly consists of star-forming galaxies rather than AGNs.

\begin{deluxetable*}{cccccccc}
\tablecaption{Our sample of [$\mathrm{O~\textsc{iii}}$] emitters at $z=3.24$ \label{table2}}
\tablehead{
\colhead{ID} & \colhead{log $M_\star$} & \colhead{log SFR} & \colhead{EW} & \colhead{log L$_\mathrm{[O~\textsc{iii}]}$} & \colhead{log L$_{1600}$} & \colhead{$\beta$} & \colhead{E(B-V)}\\
\colhead{} & \colhead{($M_\sun$)} & \colhead{($M_\sun$ yr$^{-1}$)} & \colhead{(\AA)} & \colhead{(erg s$^{-1}$)} & \colhead{(erg s$^{-1}$ Hz$^{-1}$)} & \colhead{} & \colhead{(mag)}
}
\startdata
1 & 10.21$\pm$0.24 & 1.36$\pm$0.46 & 495$\pm$174 & 42.91$\pm$0.03 & 28.42$\pm$0.02 & -1.32$\pm$0.12 & 0.26$\pm$0.09 \\
2 & 10.43$\pm$0.17 & 2.37$\pm$0.34 & 107$\pm$37 & 42.94$\pm$0.03 & 29.46$\pm$0.02 & -1.66$\pm$0.10 & 0.18$\pm$0.05 \\
3 & 9.94$\pm$0.49 & 1.29$\pm$0.44 & $>839$ & 42.83$\pm$0.16 & 28.77$\pm$0.02 &  -1.68$\pm$0.16 & 0.19$\pm$0.11 \\
4 & 9.46$\pm$0.33 & 2.26$\pm$0.19 & 173$\pm$59 & 42.92$\pm$0.03 & 28.97$\pm$0.02 & -1.82$\pm$0.10 & 0.22$\pm$0.05 \\
5 & 10.11$\pm$0.61 & 1.18$\pm$0.47 & 598$\pm$192 & 42.82$\pm$0.03 & 28.28$\pm$0.03 & -1.40$\pm$0.14 & 0.28$\pm$0.12 \\
6 & 10.15$\pm$0.18 & 1.79$\pm$0.38 & 222$\pm$70 & 42.81$\pm$0.03 & 28.81$\pm$0.02 & -1.46$\pm$0.12 & 0.23$\pm$0.07 \\
7 & 10.30$\pm$0.55 & 1.15$\pm$0.55 & $>992$ & 42.91$\pm$0.13 & 28.58$\pm$0.04 & -1.73$\pm$0.21 & 0.23$\pm$0.21 \\ 
8$\dagger$ & 9.44$\pm$0.20 & 1.05$\pm$0.30 & 592$\pm$170 & 42.85$\pm$0.02 & 28.74$\pm$0.02 & -2.20$\pm$0.11 & 0.06$\pm$0.03 \\
9$\dagger$ & 10.53$\pm$0.13 & 1.13$\pm$0.22 & 144$\pm$46 & 42.84$\pm$0.03 & 28.75$\pm$0.02 & -1.82$\pm$0.10 & 0.08$\pm$0.04 \\
10 & 9.79$\pm$0.20 & 1.23$\pm$0.36 & 342$\pm$108 & 42.81$\pm$0.03 & 28.75$\pm$0.02 & -1.99$\pm$0.10 & 0.10$\pm$0.04 \\ 
11 & 10.51$\pm$0.44 & 1.01$\pm$0.58 & 395$\pm$120 & 42.82$\pm$0.02 & 28.47$\pm$0.03 & -1.58$\pm$0.19 & 0.25$\pm$0.18 \\
12$\dagger$ & 9.47$\pm$0.18 & 1.28$\pm$0.32 & 506$\pm$169 & 42.90$\pm$0.03 & 28.87$\pm$0.02 & -2.16$\pm$0.10 & 0.08$\pm$0.04 \\
13 & 10.40$\pm$0.33 & 2.19$\pm$0.16 & 524$\pm$152 & 43.00$\pm$0.02 & 28.85$\pm$0.03 & -1.51$\pm$0.05 & 0.73$\pm$0.13 \\
14$\dagger$ & 10.72$\pm$0.17 & 2.97$\pm$0.45 & 360$\pm$119 &  42.87$\pm$0.03 & 27.83$\pm$0.02 & -1.55$\pm$0.24 & 0.68$\pm$0.12 \\ 
15 & 10.31$\pm$0.51 & 1.63$\pm$0.30 & 311$\pm$107 & 42.79$\pm$0.03 & 28.59$\pm$0.04 & -1.33$\pm$0.11 & 0.31$\pm$0.11 \\
16$\dagger$ & 9.94$\pm$0.42 & 2.62$\pm$0.10 & 265$\pm$97 & 42.79$\pm$0.03 & 29.07$\pm$0.02 &  -1.53$\pm$0.03 & 0.78$\pm$0.09 \\
17$\dagger$ & 9.42$\pm$0.33 & 0.99$\pm$0.56 & $>907$ & 42.87$\pm$0.15 & 28.42$\pm$0.03 & -1.67$\pm$0.17 & 0.37$\pm$0.28 \\
18 & 10.42$\pm$0.19 & 1.89$\pm$0.35 & 195$\pm$63 & 42.92$\pm$0.03 & 28.84$\pm$0.02 & -1.37$\pm$0.11 & 0.25$\pm$0.06 \\
19 & 9.87$\pm$0.21 & 1.73$\pm$0.37 & 534$\pm$116 & 43.14$\pm$0.02 & 29.07$\pm$0.02 & -1.83$\pm$0.12 & 0.14$\pm$0.06 \\
20& 10.56$\pm$0.41 & 2.07$\pm$0.22 & 282$\pm$80 & 42.90$\pm$0.02 & 28.92$\pm$0.02 & -1.34$\pm$0.11 & 0.44$\pm$0.12 \\ 
21$\dagger$ & 10.08$\pm$0.44 & 1.87$\pm$0.33 & 408$\pm$125 & 42.94$\pm$0.02 & 28.80$\pm$0.03 & -1.41$\pm$0.10 & 0.31$\pm$0.11 \\
22 & 9.82$\pm$0.17 & 2.40$\pm$0.21 & 239$\pm$64 & 42.94$\pm$0.02 & 29.01$\pm$0.03 & -1.50$\pm$0.08 & 0.27$\pm$0.04 \\
23 & 9.92$\pm$0.13 & 2.88$\pm$0.06 & 281$\pm$107 & 42.83$\pm$0.03 & 29.27$\pm$0.02 & -1.55$\pm$0.01 & 0.85$\pm$0.05 \\
24 & 9.91$\pm$0.20 & 2.36$\pm$0.27 & 479$\pm$145 & 42.89$\pm$0.02 & 28.55$\pm$0.02 & -0.97$\pm$0.16 & 0.36$\pm$0.07 \\
\enddata
\tablecomments{Protocluster members are denoted by the $\dagger$ behind their IDs. Three galaxies are not detected in the continuum, therefore we use the 2$\sigma$ flux limit to give a lower limit of the rest-frame EW. The [$\mathrm{O~\textsc{iii}}$] luminosity L$_\mathrm{[O~\textsc{iii}]}$ are observed values (not corrected for dust).}
\end{deluxetable*}

\section{A large galaxy overdensity traced by [$\mathrm{O~\textsc{iii}}$] emitters} \label{sec4}

\subsection{The spatial distribution of galaxies}  \label{distribution}
In \cite{Shi21}, we identified a large galaxy overdensity traced by photo-$z$ galaxies using Voronoi tessellation technique. We refer the readers to \cite{Shi21} for further details about the method used to obtain the surface density map. Briefly speaking, a Voronoi tessellation is a scale-independent method which devides the 2D distribution of points into convex cells, with the local density $f$ of each cell being the inverse of the cell area. Therefore, smaller cells correspond to denser regions. The density contrast ($\tilde{f}$) of each point is defined as $\tilde{f}=f/\langle f \rangle$, where $\langle f \rangle$ is the average density in the entire field.

\begin{figure*}[ht!]
\plotone{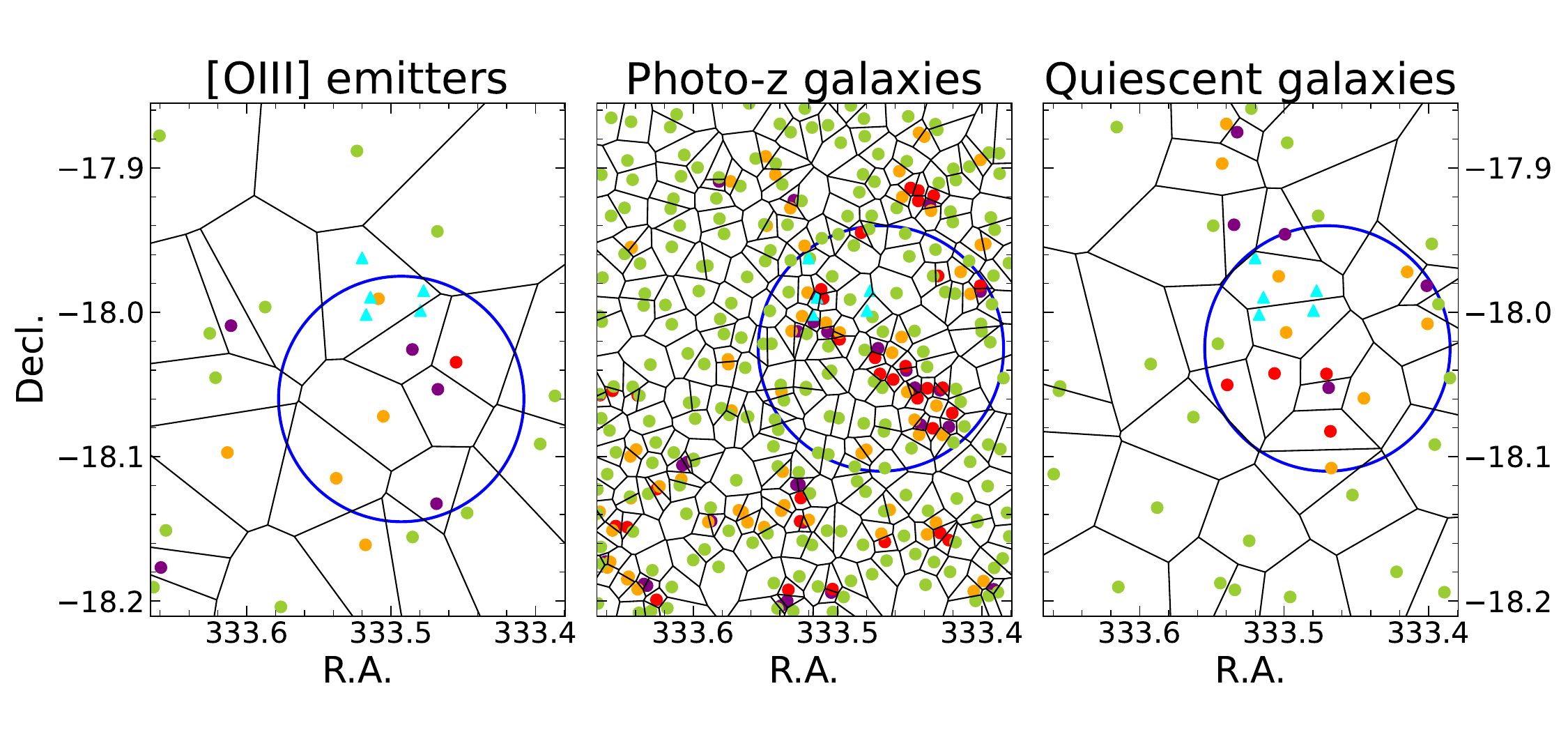}
\epsscale{1.2}
\caption{Voronoi tessellation of [$\mathrm{O~\textsc{iii}}$] emitters (left), photo-$z$ galaxies (middle) and quiescent galaxies (right). The different colored points respresent galaxies with different local density contrast: $\tilde{f}>2.0$ (purple), $1.5<\tilde{f}<2.0$ (red), $1.0<\tilde{f}<1.5$ (orange) and $\tilde{f}<1.0$ (green). The cyan triangles are the five spectroscopically confirmed LBGs at z = 3.24. The large blue circles (10 Mpc in radius) denote the protocluster regions traced by different galaxy populations. }

\label{figure4}
\end{figure*}

The spatial distribution of our [$\mathrm{O~\textsc{iii}}$] emitters is shown in the left panel of Figure \ref{figure4}. A clear overdensity can be seen in the middle of the field. As in \cite{Shi21}, we also use a 10 Mpc radius circle to denote the protocluster region, which is consistent with the typical size of protoclusters at $z\sim3$ \citep{Chiang13}. To include as many protocluster members as possible, the circle is placed to include four adjacent spectroscopically confirmed LBGs \citep{Toshikawa16} as well as most of the [$\mathrm{O~\textsc{iii}}$] emitters in the overdense regions (i.e., $\tilde{f}>1.0$). 

For comparison, in Figure \ref{figure4} we also show the distributions of photo-$z$ and quiescent galaxies selected in \cite{Shi21}. In these cases, the protocluster region is also denoted by a 10 Mpc radius circle enclosing the largest photo-$z$ overdensity in the field as well as all the five spectroscopic sources. The protocluster region traced by 
[$\mathrm{O~\textsc{iii}}$] emitters largely coincide with that traced by photo-$z$ and quiescent galaxies: the spatial offset between the centers of the circles is smaller than 2$\arcmin$. However, we note that the overdensity traced by [$\mathrm{O~\textsc{iii}}$] emitters appears to be skewed toward the southeast. Three galaxies in high-density regions outside the circle are located in the smaller overdense regions of photo-$z$ galaxies in the southeast. On the other hand, the overdensity traced by quiescent galaxies tend to be located toward the northwest and extend further north, which is also seen in the photo-$z$ galaxy distribution. 

Figure \ref{figure4} implies that there may be a spatial segregation between star-forming galaxy population such as [$\mathrm{O~\textsc{iii}}$] emitters and quiescent galaxy population, which further suggests that they may trace different underlying large-scale structures. We will discuss this phenomenon in detail in Section \ref{discuss2}.

\subsection{Descendant mass of D4UD01}
The narrow-band selected [$\mathrm{O~\textsc{iii}}$] emitters render us an opportunity to estimate the intrinsic overdensity of D4UD01 and infer its present-day mass, since they are minimally contaminated by fore- or background interlopers, and are selected in a consistent way. In this section, we discuss the future fate of this protocluster in detail.

There are seven [$\mathrm{O~\textsc{iii}}$] emitters in the 10 Mpc radius protocluster region (blue circle in Figure \ref{figure4}). The area of the circle is 74 arcmin$^2$, which results in a surface density of $\Sigma=0.09\pm 0.03$ arcmin$^{-2}$, where the error reflects the Poisson noise. To calculate the average density of [$\mathrm{O~\textsc{iii}}$] emitters in the field, we exclude the protocluster area to avoid a strong bias toward higher density values. The remaining 17 [$\mathrm{O~\textsc{iii}}$] emitters are distributed in the 302 arcmin$^2$ area, resulting in an average surface density of $\overline{\Sigma}=0.05\pm 0.01$ arcmin$^{-2}$. Therefore, the overdensity of the protocluter region is $\delta_\Sigma=(\Sigma-\overline{\Sigma})/\overline{\Sigma}=0.8\pm0.6$. In comparison, the overdensity of photo-$z$ galaxies shown in Figure \ref{figure4} is $\delta_\Sigma=0.5\pm0.2$ and the overdensity of quiescent galaxies is $\delta_\Sigma=1.0\pm0.5$. Thus our [$\mathrm{O~\textsc{iii}}$] overdensity is even higher than the photo-$z$ overdensity and comparable to that of quiescent galaxies. 

Based on Millennium Simulations \citep{Springel05}, \cite{Chiang13} estimated the probability of a structure being a genuine protocluster given the galaxy overdensity $\delta_g$. For typical star-forming galaxies, using a cubic box of (25 Mpc)$^3$ (similar to the volumn occupied by the [$\mathrm{O~\textsc{iii}}$] overdensity), they found that a structure with an overdensity of $\delta_g=0.64$ has a 50\% probability of evolving into a galaxy cluster by $z=0$, increasing to 80\% confidence for $\delta_g=1.10$. The overdensity of our protocluster is 0.8, falling between the 50\% and 80\% confidence level, which further supports its authenticity.
In addition, \cite{Chiang13} also showed the galaxy overdensities as a function of present-day cluster mass at different redshfts. According to their Figure 6, protoclusters with overdensities of $\delta_g=0.6-1.2$ at $z\sim3$ will evolve into ``Fornax-type'' clusters of $1.37-3 \times 10^{14} M_\odot$ by present day. It should be noted that this mass estimate is only a lower limit, since the line-of-sight distance sampled by $H_2S1$ band is $\sim$2 times larger than their window size. As demonstrated in \cite{Chiang13}, the significance of the observed overdensity usually declines rapidly with increasing redshift uncertainty $\Delta z$, due to the projection of low-density interlopers. Lastly, we note that \cite{Toshikawa16} also estimated the descendant halo mass of this protocluster traced by LBGs using lightcone models \citep{Henriques12} from Millennium Simulation. Their estimated mass, $1.5-5 \times 10^{14} M_\odot$, aligns with the values inferred by \cite{Chiang13}.

Next, we adopt an analytic approach to calculate the descendant mass of the protocluster as in \cite{Steidel98}. Assuming all the enclosed mass within the protocluster region becomes virialized by $z=0$, then the total mass of the protocluster can be expressed as:
\begin{equation}
M_{z=0}=(1+\delta_m)~\overline{\rho}~V_{true}
\end{equation}
where $\overline{\rho}$ is the average matter density of the universe. The matter overdensity $\delta_m$ is related to galaxy overdensity as $1+b \delta_m = C (1+\delta_g)$ where $b$ is the bias parameter, and $C$ is the correction factor accounting for the effect of redshift space distortion due to peculiar velocities. $V_{true}$ is the true volume of the protocluster which is related to the observed volume as $V_{true}=V_{obs}/C$. In the case of spherical collapse model, $C=1+\Omega^{4/7}(z)[1-(1+\delta_m)^{1/3}]$. Our protocluster has a galaxy overdensity of $0.8\pm0.6$ with a volume of $\sim 1.7 \times 10^4$ Mpc$^3$. Adopting a bias value of $b=3.4$ for [$\mathrm{O~\textsc{iii}}$] emitters at $z=3$ from \cite{Zhai21}, we obtain a matter overdensity of $\delta_m=0.2\pm0.1$. Thus the total mass of the protocluster is $M_{z=0}=(8.5\pm0.7) \times 10^{14} M_\odot$. The mass calculated in this method is mainly determined by the volume and the mass overdensity. Usually, using a larger volume would diminish the enclosed overdensity, and vice versa; however their product should not change much. To verify this, we define two circular regions with radii of 8 Mpc and 12 Mpc, each encompassing only the 7 protocluster galaxies, as the protocluster regions. By repeating the above calculations, we estimate the total masses as $M_{z=0}=(6.6\pm1.2) \times 10^{14} M_\odot$ and $M_{z=0}=(9.1\pm0.8) \times 10^{14} M_\odot$, respectively. These are consistent with our previous estimate within uncertainties. Thus we conclude that the overdensity traced by  [$\mathrm{O~\textsc{iii}}$] emitters will most likely evolve into a Virgo-like ($3-10 \times 10^{14} M_\odot$) cluster by the present day.

\section{Discussion} \label{sec5}
\subsection{Environmental Impacts on Physical Properties of Galaxies} \label{env}
After identifying the protocluster in the field, in this section we aim to compare the physical properties of [$\mathrm{O~\textsc{iii}}$] emitters in and out of the protocluster, to further discern possible environmental impacts on galaxy properties. Since our sample is selected in a uniform way, it minimizes selection bias and provides a more reliable approach for evaluating environmental effects. 

\begin{figure}[ht!]
\plotone{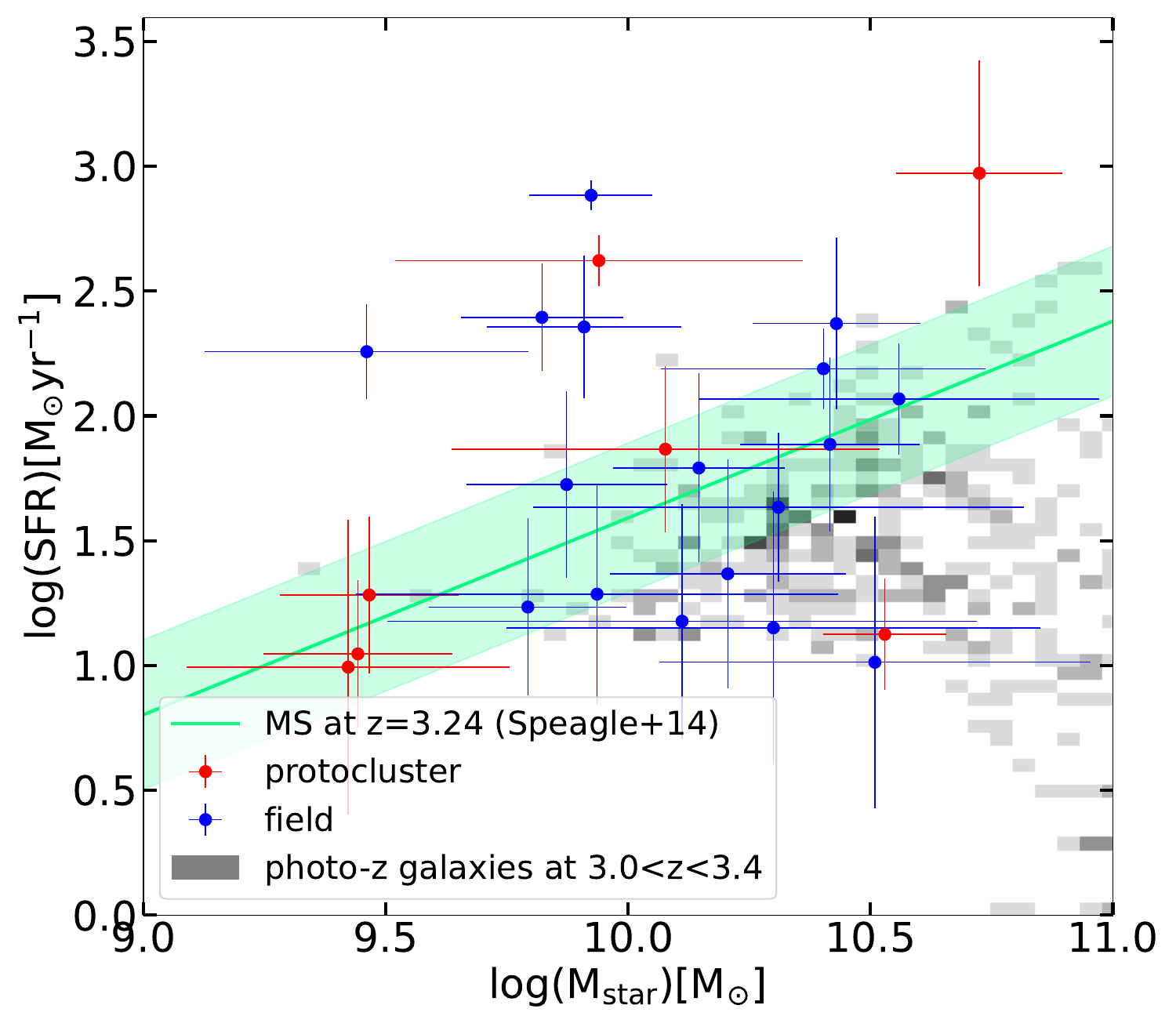}
\caption{
  SFR-M$_{\star}$ relation for the [$\mathrm{O~\textsc{iii}}$] emitters in our sample. The star-forming main sequence (MS) at $z=3.24$ derived from \cite{Speagle14} is indicated as the green solid line with a 0.3 dex dispersion. The grey shaded area shows the $K_S$ band selected photo-$z$ galaxies at $3.0<z<3.4$ from \cite{Shi21}.
}
\label{figure5}
\end{figure}

It is well known that star-forming galaxies generally follow a tight ``main-sequence'' (MS) on the SFR-M$_{\star}$ plane, namely ``star-forming main sequence'' \citep[e.g.,][]{Daddi07,Elbaz07,Noeske07,Koyama13}. Figure \ref{figure5} shows our [$\mathrm{O~\textsc{iii}}$] emitters on the SFR-M$_{\star}$ plane in different environments. Accounting for the errorbars, most of the [$\mathrm{O~\textsc{iii}}$] emitters follow the star-forming main sequence at $z=3.24$ within 0.3 dex with only a few exceptions. This further suggests that these [$\mathrm{O~\textsc{iii}}$] emitters are predominantly normal star-forming galaxies driven by secular star-formation activities. Similar results have also been reported in the literature \citep[e.g.,][]{Suzuki15,Onodera16}. However, we also note that several studies have found the SFRs of the [$\mathrm{O~\textsc{iii}}$] emitters appear to be elevated relative to the MS \citep{Onodera20,Wen22}. In some cases the enhanced star-formation activities of these emission line galaxies could be attributed to the associated protocluster environments \citep{Wen22}. It is also possible that the low-mass emission line galaxies are undergoing a starburst phase as seen in \cite{Onodera20}, where they found the low-mass ($M_\star<10^{9.5}$ M$_\sun$) [$\mathrm{O~\textsc{iii}}$] emitters are located above the MS, while high mass ($M_\star>10^{9.5}$ M$_\sun$) ones remain on the MS. This phenomenon has also been observed for other emission line galaxies such as H$\alpha$ emitters \citep{Hayashi16}. In comparison, due to our survey's detection limits, our sample mainly consists of high-mass ($M_\star>10^{9.5}$ M$_\sun$) galaxies, which likely represent more evolved populations that have already progressed beyond their starburst stage. Therefore, we do not observe a significant enhancement in star formation activity across our sample.

\begin{figure*}[ht!]
\plotone{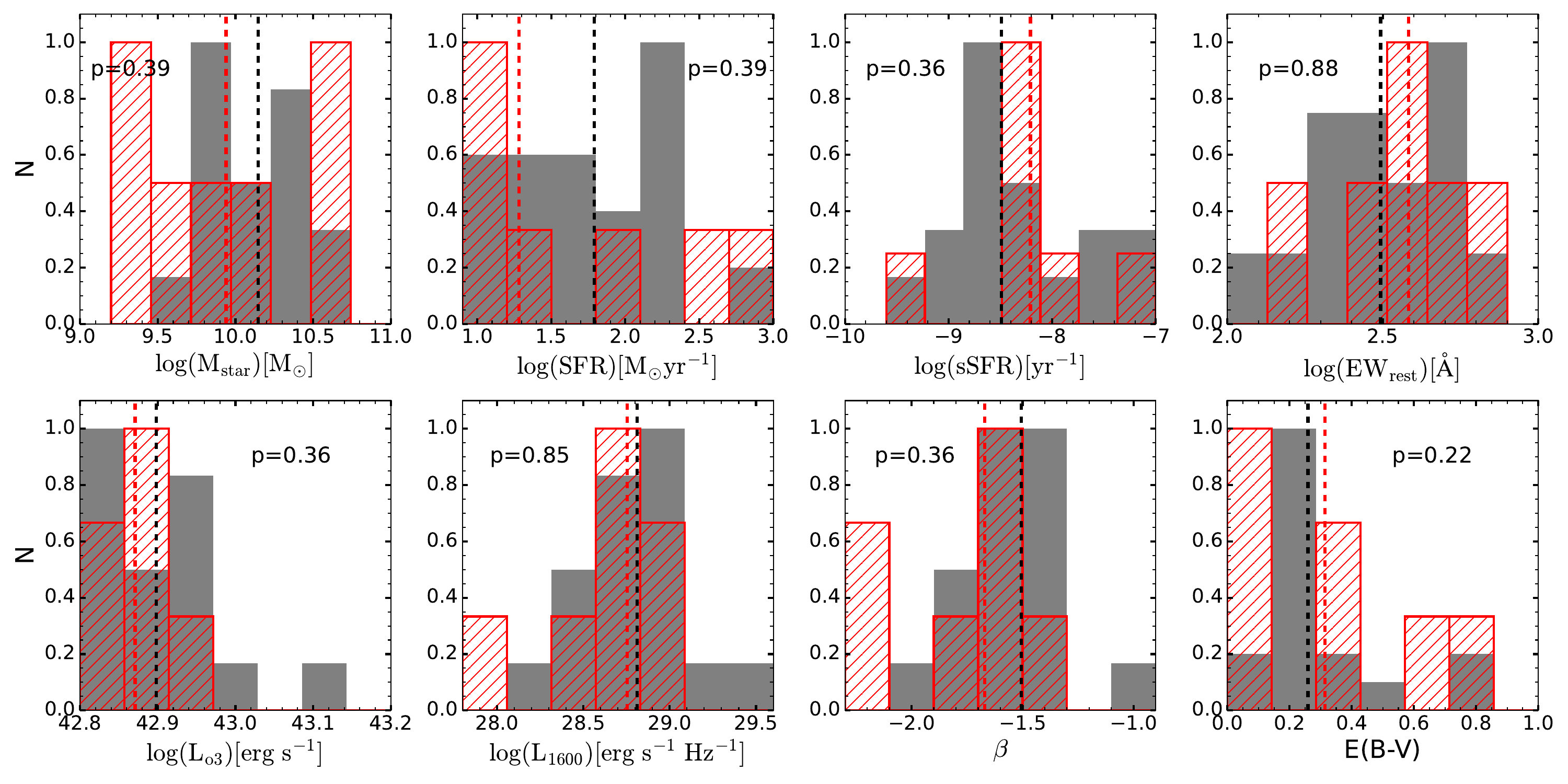}
\caption{
  Normalized distributions of the physical properties of [$\mathrm{O~\textsc{iii}}$] emitters in the protocluster (red hatched) and in the field (gray). The corresponding $p$-value obtained from K-S test is indicated in each panel. The red and black vertical lines represent the median values of the protocluster and field galaxies, respectively.
}
\label{figure6}
\end{figure*}

Next, we designate 7 galaxies within the protocluster region as protocluster members, while classifying the remaining 17 as field galaxies. For the two subsamples, we compare their physical properties in Figure \ref{figure6}. We show the distributions of the physical parameters as listed in Table \ref{table2}. To evaluate the differences between the two subsamples, we perform the K-S test and give the $p$-values in the figure as well. Since our choice of a 10-Mpc radius for defining protocluster region is somewhat arbitrary, we also perform a sanity check using a 7-Mpc radius circle containing 5 protocluster galaxies and a 13-Mpc circle enclosing 10 protocluster members to redo the above analysis, and our following results remain consistent across all three radius choices, suggesting our results are not sensitive to the specific choice of protocluster boundary.

Figure \ref{figure6} reveals no significant differences in physical properties between the two subgroups, as can be clearly inferred from the large $p$-values.
Although field galaxies show higher median stellar masses and SFRs compared to protocluster galaxies, their overall distributions are statistically indistinguishable (also see Figure \ref{figure5}). The same is true for specific SFR (sSFR$=$SFR/M$_\star$): no significant difference can be spotted despite the protocluster members exhibit higher median values. Similarly, the distributions of rest-frame EWs, [$\mathrm{O~\textsc{iii}}$] luminosities, UV luminosities, UV slope and dust attenuations are all indistinguishable from the field. These similar physical properties suggest that protocluster galaxies exhibit comparable evolutionary behavior to field galaxies. 

Since our sample size is small, the K-S test could become less sensitive to differences in distributions that may fail to reject the null hypothesis when it is false. To validate the K-S test results, we implemented a non-parametric permutation test as a complementary method. To do this, we calculated the original K-S statistic, then generated 1000 permutations by randomly reassigning observations between the two subsamples while maintaining the original sample sizes. The empirical $p$-value was determined as the proportion of permuted statistics that equaled or exceeded our observed value. This distribution-free approach makes no assumptions about asymptotic behavior of the original K-S test and derives the sampling distribution directly from our observed data, making it particularly suitable for our small sample size. The permutation-derived $p$-values were all very close to the original K-S test values (maximum difference of 0.02), thus strengthening confidence in our distributional comparison results.

Next we discuss the evolutionary state of D4UD01. \cite{Shi21} found an elevated fraction of quiescent galaxies in D4UD01, suggesting it may represent an evolved structure.  Evolved protoclusters are expected to contain a significant population of quiescent galaxies. While identifying quiescent galaxies at high redshifts remains challenging, several recent studies have successfully detected multiple high-z ($z > 2$) protoclusters with substantial populations of spectroscopically confirmed quiescent galaxies \citep{McConachie22,Ito23, Shi24b, Tanaka24, Jin24,Naufal24}. These observations vary in survey depth, potentially missing many low-mass quiescent galaxies and complicating the interpretation of quiescent fractions across protoclusters. To ensure a mass complete sample, we apply a mass cut of $M_\star>10^{11}M_\sun$ to some of these observations and calculate their quiescent fractions.

\begin{figure}[ht!]
\plotone{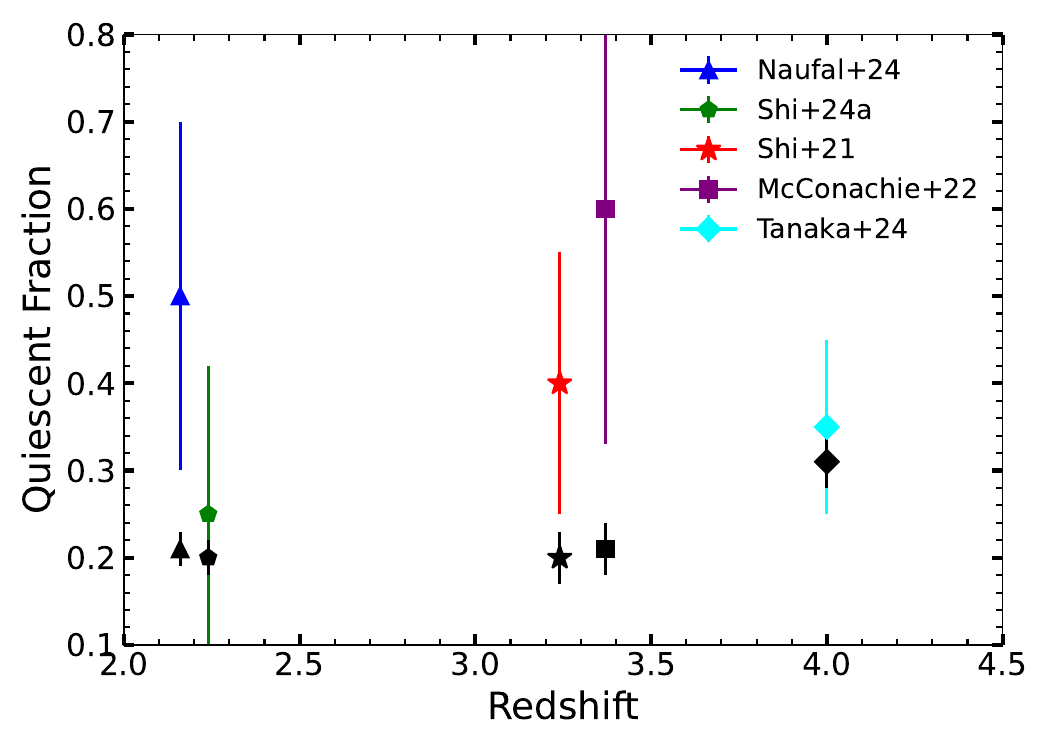}
\caption{
  The quiescent fraction of different protoclusters at different redshifts. A mass-cut of $M_\star>10^{11}M_\sun$ is adopted when selecting quiescent galaxies. From left to right, protoclusters are located at $z=2.16$ \citep{Naufal24}, $z=2.24$ \citep{Shi24b}, $z=3.24$ \citep[our protocluster,][]{Shi21}, $z=3.37$ \citep{McConachie22} and $z=4$ \citep{Tanaka24}. The colored symbols on the top represent the corresponding quiescent fractions. The errorbars are computed using Poisson statistics in cases where uncertainties are not provided in the relevant papers. The black symbols on the bottom denote the quiescent fractions of the COSMOS field at the protocluster redshifts from \cite{Weaver23}.
}
\label{figure7}
\end{figure}

We compare the quiescent fraction of our protocluster ($\sim$40\% as in \cite{Shi21}) with these studies in Figure \ref{figure7}. The quiescent fraction varies in different protoclusters, reflecting their different evolutionary states. To make a consistent comparison, we also plot the quiescent fraction of field galaxies in the COSMOS field \citep{Weaver23} at each protocluster redshift. Notably, the Spiderweb protocluster at $z=2.16$ \citep{Naufal24} and MAGAZ3NE J0959 protocluster \citep{McConachie22} at $z=3.37$ stand out as having the highest fraction of quiescent galaxies, suggesting they probably have already entered a maturing phase. In comparison, the BOSS1244 protocluster at $z=2.24$ \citep{Shi24b} and the protocluster at $z=4$ discovered by \cite{Tanaka24} show quiescent fraction similar to the field, indicating they are likely still at their early evolutionary stages. Our protocluster has the quiescent fraction two times higher than the field, also suggesting that it is entering a late evolutionary phase.   This finding demonstrates that protoclusters at similar redshifts can be in markedly different evolutionary states.

However, if our protocluster is at a later evolutionary phase, it seems to conflict with the finding of this work, that our [$\mathrm{O~\textsc{iii}}$] emitters show no environmental dependence in their properties. We propose that this apparent contradiction could be reconciled by considering the timescale of quenching in the protocluster environment. If the transition from active star formation to quiescence occurs in a short time scale, we would observe two major populations: actively star-forming galaxies that have not yet begun quenching (our [$\mathrm{O~\textsc{iii}}$] emitters) and already-quenched galaxies (as discovered in \cite{Shi21}), with few galaxies caught in the intermediate phase such as ``Green Valley'' galaxies \citep{Salim14,Angthopo19}. This rapid-quenching scenario would explain why we do not detect many ``fading'' [$\mathrm{O~\textsc{iii}}$] emitters with diminished star formation rates in the protocluster. Future observations targeting galaxies in various stages of quenching could help validate this scenario and better constrain the star formation histories of protocluster galaxies.

The diversity of environmental effects observed in different protoclusters could be attributed to their varying stages of evolution. Protoclusters exhibiting enhanced star formation activity relative to the field population typically represent systems in their early evolutionary stages, when gas-rich environments and frequent galaxy interactions drive elevated star formation. Conversely, protoclusters showing suppressed star formation and an enhanced quiescent fraction likely represent more evolved systems, where mechanisms such as gas depletion, environmental quenching, and possibly early mass assembly have already begun to dampen star formation activity. This evolutionary sequence suggests that protoclusters, despite their similar final fates as massive clusters, can present substantially different properties depending on their stage of development at the observed epoch.

Alternatively, it is also possible that our sample size is too small to discern any environmental trends. \cite{Cucciati14} and \cite{Lemaux18} reported the discovery of two protoclusters at $z=2.9$ and $z=4.57$ respectively, and they did not find any differences in the star-formation activities between protocluster and field galaxies. While these studies used more precise spectroscopic data, their small sample sizes (fewer than 10 protocluster galaxies) may have limited their statistical power to detect environmental effects. The consistent lack of detected environmental effects across these studies could reflect either a genuine absence of environmental influence at these redshifts or, more likely, insufficient statistical power to detect subtle environmental trends. Larger surveys with expanded protocluster galaxy samples, combined with deep spectroscopic observations, will be essential for definitively resolving this question.

Last but not least, far-IR and submilimeter observations are further needed to perform a comprehensive census of this protocluster. In lack of observations beyond IRAC $4.5\micron$, rest-frame infrared information that capture dust content and polycyclic aromatic hydrocarbons would be missing in our [$\mathrm{O~\textsc{iii}}$] and photo-$z$ samples. These infrared emissions are generally excited by intense star-formation or AGNs. Therefore, to confirm the ``red and dead'' nature of the quiescent galaxies in this protocluster, future submillimeter observations with ALMA  will be essential to provide crucial constraints on any residual star formation activity that may be obscured in optical wavelengths.

\subsection{Diverse types of galaxies in the protocluster field} \label{discuss2}
In this work, we identify 24 narrow-band selected [$\mathrm{O~\textsc{iii}}$] emitting galaxies in the field of D4UD01, which represent normal star-forming galaxies. In our previous work \citep{Shi21}, we also identified 450 galaxies with photometric redshift $3.0<z<3.4$ in the field, among which 52 are classified as quiescent galaxies. These samples showcase diverse types of galaxy populations tracing this protocluster (Figure \ref{figure4}).

\begin{figure*}[ht!]
\plotone{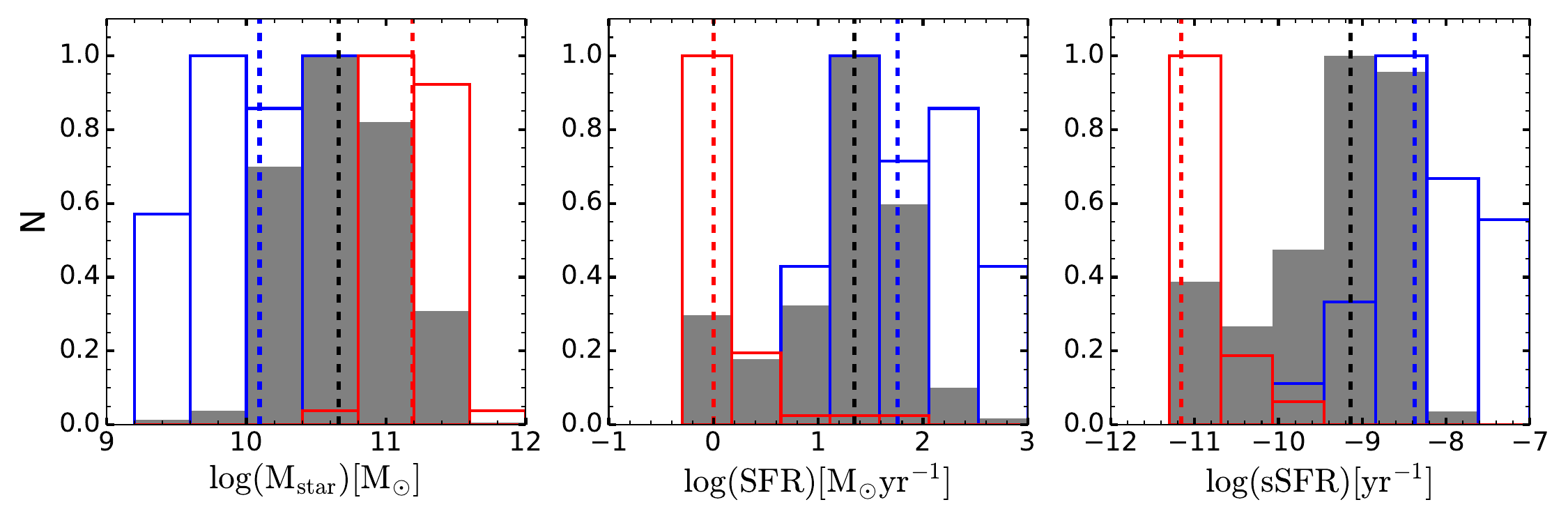}
\caption{
  Normalized distributions of stellar mass (left), star-formation rate (middle) and specific star-formation rate (right) of [$\mathrm{O~\textsc{iii}}$] emitters (blue), photo-$z$ galaxies (gray) and photo-z selected quiescent galaxies (red). Galaxies with SFR$=0$ are are indicated in the log(SFR)$=0$ location. The vertical lines indicate the median values of each distribution.
}
\label{figure8}
\end{figure*}

Figure \ref{figure8} shows the distribution of stellar mass, SFR and sSFR values measured in each sample. It is obvious that the [$\mathrm{O~\textsc{iii}}$] emitters are low mass ($M_\star\sim10^{10}$ M$_\sun$) star-forming galaxies with much higher sSFR than photo-$z$ galaxies, especially quiescent galaxies. A K-S test between [$\mathrm{O~\textsc{iii}}$] emitters and photo-$z$ galaxies (also quiescent galaxies) clearly shows they are drawn from distinct distributions ($p$-value $\ll 0.01$ for all the three parameters). Since our photo-$z$ sample is incomplete below $M_\star\sim10^{10.8}$ M$_\sun$ \citep{Shi21}, [$\mathrm{O~\textsc{iii}}$] emitters probe the low-mass end of galaxy stellar mass function, enabling a more comprehensive investigation of the protocluster's extent.

In Section \ref{distribution}, we see that there appears to be a spatial offset between [$\mathrm{O~\textsc{iii}}$] emitters and photo-$z$ quiescent galaxies. We further quantify their spatial correlation using the angular two-point cross-correlation function (CCF). The angular two-point correlation function is a way to describe the excess probability of finding two galaxies separated by a certain angular distance, relative to the random distributions. This technique has been utilized in the literature to investigate the spatial association between different galaxy populations \citep[e.g.,][]{Tamura09,Harikane19,Shi20}. We calculate the CCF using the \cite{Landy93} estimator:
\begin{equation}
\omega(\theta)=\frac{D_1D_2(\theta)-D_1R_2(\theta)-R_1D_2(\theta)+R_1R_2(\theta)}{R_1R_2(\theta)},
\end{equation}
where $DD$, $DR$, $RD$, and $RR$ are the galaxy-galaxy, galaxy-random, random-galaxy, and random-random pair counts, respectively, for groups 1 and 2. The statistical errors of the CCFs are estimated from the standard deviation of 1000 bootstrap realizations.

\begin{figure}[ht!]
\plotone{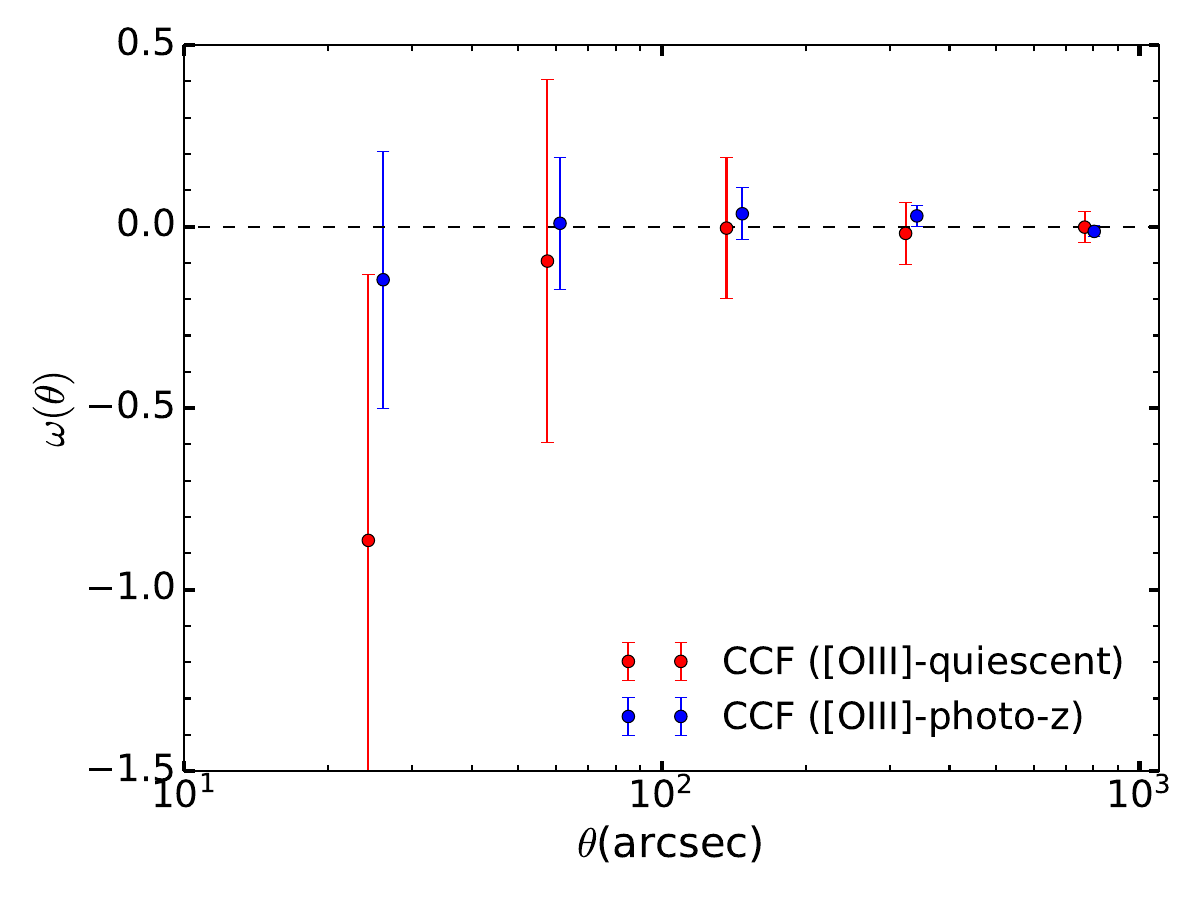}
\caption{
Angular cross-correlation between different galaxy populations with the [$\mathrm{O~\textsc{iii}}$] emitters. The blue, red circles show the CCFs between photo-$z$ galaxies and [$\mathrm{O~\textsc{iii}}$] emitters, quiescent galaxies and [$\mathrm{O~\textsc{iii}}$] emitters, respectively. The errors are estimated from the standard deviation of 1000 bootstrap samples.
}
\label{figure9}
\end{figure}

Figure \ref{figure9} shows the CCFs between the [$\mathrm{O~\textsc{iii}}$] emitters and the photo-$z$ galaxies as well as the photo-$z$ quiescent galaxies. At angular scales below $\sim1\arcmin$, we observe a strong anti-correlation between [$\mathrm{O~\textsc{iii}}$] emitters and quiescent galaxies. A mild anti-correlation is also seen for photo-$z$ galaxies. This further confirms our visual impression in Figure \ref{figure4}, where we see that the overdensity of [$\mathrm{O~\textsc{iii}}$] emitters is skewed towards southeast relative to the overdensities of photo-$z$ and quiescent galaxies.

The discrepancy in spatial distribution between different galaxy populations is frequently reported in protocluster studies \citep[e.g.,][]{Shi191,Shi192,Shi20,Ito21,Yonekura22,Zhang22}. For example, \cite{Shi191} conducted a multi-wavelength study of a protocluster at $z=3.78$, finding a strong disparity between the spatial distribution of Balmer break galaxies and Lyman-alpha emitters (LAEs). They argued that those Balmer break galaxies may be located in the center of the most massive dark matter halos while Lyman-alpha emitters prefer to reside in low-mass halos. The different formation times of the underlying dark matter halos lead to the well-known halo assembly bias \citep[e.g.,][]{Gao05,Wechsler06,Li08,Zentner14}. \cite{Ito21} also investigated the spatial distributions of LAEs and massive star-forming galaxies and quiescent galaxies at $2<z<4.5$, they found that LAEs tend to reside in less dense regions compared to other star-forming galaxies and quiescent galaxies, in line with \cite{Shi191}. Similarly, \cite{Yonekura22} studied a protocluster at $z=2.39$ and found quiescent galaxies tend to avoid the density peak of Lyman-alpha emitters. These studies suggest that different galaxy populations may trace different dark matter halos that formed at different epochs. 

In the above cases, old massive quiescent galaxies might assemble first in the central dense region of the protocluster, followed by the formation of star-forming galaxies in the outskirts, in accordance with the ``inside-out'' scenerio \citep{Chiang17}. This scenario was also corroborated by recent observational studies. In particular, \cite{Shi24b} spectroscopically confirmed two massive quiescent galaxies in the BOSS1244 protocluster at $z=2.24$, which are located in the center of the protocluster, surrounded by star-forming galaxies in the outskirts. It is argued that these two quiescent galaxies are merging into a (proto-)brightest cluster galaxy even before virialization.
The spatial segregation between our [$\mathrm{O~\textsc{iii}}$] emitters and quiescent galaxies in this study further supports this scenerio. In this context, the photo-$z$ overdensity may trace a more evolved halo where many quiescent galaxies can be found; while the [$\mathrm{O~\textsc{iii}}$] overdensity represents a more recently formed structure that host normal star-forming galaxies. Future extensive spectroscopy could elucidate the true configurations of these structures.
Therefore, an important implication from these findings is that, one should exercise more caution when attempting to use a single galaxy population to study protoclusters, as it may be a biased tracer of the underlying large-scale structure.

\section{Summary and Conclusion} \label{sum}
In this paper, we investigate a protocluster at $z=3.24$ in the CFHTLS D4 field (D4UD01) with [$\mathrm{O~\textsc{iii}}$] emitters obtained from narrow-band observations. Using the deep narrow-band $H_2S1$ filter combined with the broad-band $K_S$ image from CFHTLS, we identify a robust sample of 24 [$\mathrm{O~\textsc{iii}}$] emitters at $z=3.24$. With multiwavelength data, we perform SED fitting to derive their physical properties to study the environmental effects in detail. Our major results are summarized below.

1. Our [$\mathrm{O~\textsc{iii}}$] emitters exhibit strong emission lines with a median value rest-frame equivalent width of 342\AA. Galaxies in our sample have a median mass of $M_\star\sim10^{10}$ M$_\sun$, and are all star-forming galaxies mostly following the main sequence of star-formation. 

2. The [$\mathrm{O~\textsc{iii}}$] emitters form a large overdensity ($\delta=0.8$) in the field, which largely coincides with the photo-$z$ galaxy overdensity identified in \cite{Shi21}. We estimate the total mass enclosed in the overdensity to be $\sim 8.5 \times 10^{14} M_\odot$, suggesting it would evolve into a Virgo-like cluster by the present day.

3. The [$\mathrm{O~\textsc{iii}}$] emitters are distinct galaxy populations from photo-$z$ galaxies in the same field, in a sense that they are less massive and have higher star-formation rates. There is also a spatial offset between the concentrations of [$\mathrm{O~\textsc{iii}}$] emitters and that of the photo-$z$/quiescent galaxies. The angular correlation function reveals a strong anti-correlation between [$\mathrm{O~\textsc{iii}}$] emitters and quiescent galaxies at small scales. One possible interpretation is that the less massive [$\mathrm{O~\textsc{iii}}$] emitters and massive photo-$z$/quiescent galaxies inhabit different dark matter halos that formed at different epochs: more recent for the former and earlier for the latter. This ``halo-assembly bias'' indicates that relying on a single galaxy population may provide an incomplete view of the protocluster structures.

4. The physical properties of [$\mathrm{O~\textsc{iii}}$] emitters show no significant differences between protocluster and field environments. Combined with the high fraction of quiescent galaxies in this protocluster, it suggests that this protocluster may represent a more evolved structure that has progressed beyond its peak star-formation phase, although our limited sample size may restrict our ability to detect subtle environmental effects. We argue that the different environmental effects found in different protoclusters may imply that they are in different evolutionary states.

\section*{Acknowledgments}
We thank the anonymous referee for a careful review and helpful comments that improved this paper. K.S. acknowledges the funding from Fundamental Research Funds for the Central Universities under Grant No. SWU-KR22035. X.Z.Z. thanks supported from the National Science Foundation of China (12233005, 12073078). D.D.S. acknowledges support from the National Science Foundation of China (Grant No. 12303015), the National Science Foundation of Jiangsu Province (Grant No. BK20231106), and the China Manned Space Project. This research adopts data obtained through the Telescope Access Program (TAP), which is funded by the National Astronomical Observatories and the Special Fund for Astronomy from the Ministry of Finance.

\bibliography{paper}

\begin{thebibliography}{}
\expandafter\ifx\csname natexlab\endcsname\relax\def\natexlab#1{#1}\fi
\providecommand{\url}[1]{\href{#1}{#1}}
\providecommand{\dodoi}[1]{doi:~\href{http://doi.org/#1}{\nolinkurl{#1}}}
\providecommand{\doeprint}[1]{\href{http://ascl.net/#1}{\nolinkurl{http://ascl.net/#1}}}
\providecommand{\doarXiv}[1]{\href{https://arxiv.org/abs/#1}{\nolinkurl{https://arxiv.org/abs/#1}}}

\bibitem[{{Abdurro'uf} {et~al.}(2022){Abdurro'uf}, {Accetta}, {Aerts}, {Silva
  Aguirre}, {Ahumada}, {Ajgaonkar}, {Filiz Ak}, {Alam}, {Allende Prieto},
  {Almeida}, {Anders}, {Anderson}, {Andrews}, {Anguiano}, {Aquino-Ort{\'\i}z},
  {Arag{\'o}n-Salamanca}, {Argudo-Fern{\'a}ndez}, {Ata}, {Aubert},
  {Avila-Reese}, {Badenes}, {Barb{\'a}}, {Barger}, {Barrera-Ballesteros},
  {Beaton}, {Beers}, {Belfiore}, {Bender}, {Bernardi}, {Bershady}, {Beutler},
  {Bidin}, {Bird}, {Bizyaev}, {Blanc}, {Blanton}, {Boardman}, {Bolton},
  {Boquien}, {Borissova}, {Bovy}, {Brandt}, {Brown}, {Brownstein}, {Brusa},
  {Buchner}, {Bundy}, {Burchett}, {Bureau}, {Burgasser}, {Cabang}, {Campbell},
  {Cappellari}, {Carlberg}, {Wanderley}, {Carrera}, {Cash}, {Chen}, {Chen},
  {Cherinka}, {Chiappini}, {Choi}, {Chojnowski}, {Chung}, {Clerc}, {Cohen},
  {Comerford}, {Comparat}, {da Costa}, {Covey}, {Crane}, {Cruz-Gonzalez},
  {Culhane}, {Cunha}, {Dai}, {Damke}, {Darling}, {Davidson}, {Davies},
  {Dawson}, {De Lee}, {Diamond-Stanic}, {Cano-D{\'\i}az}, {S{\'a}nchez},
  {Donor}, {Duckworth}, {Dwelly}, {Eisenstein}, {Elsworth}, {Emsellem},
  {Eracleous}, {Escoffier}, {Fan}, {Farr}, {Feng}, {Fern{\'a}ndez-Trincado},
  {Feuillet}, {Filipp}, {Fillingham}, {Frinchaboy}, {Fromenteau}, {Galbany},
  {Garc{\'\i}a}, {Garc{\'\i}a-Hern{\'a}ndez}, {Ge}, {Geisler}, {Gelfand},
  {G{\'e}ron}, {Gibson}, {Goddy}, {Godoy-Rivera}, {Grabowski}, {Green},
  {Greener}, {Grier}, {Griffith}, {Guo}, {Guy}, {Hadjara}, {Harding},
  {Hasselquist}, {Hayes}, {Hearty}, {Hern{\'a}ndez}, {Hill}, {Hogg},
  {Holtzman}, {Horta}, {Hsieh}, {Hsu}, {Hsu}, {Huber}, {Huertas-Company},
  {Hutchinson}, {Hwang}, {Ibarra-Medel}, {Chitham}, {Ilha}, {Imig}, {Jaekle},
  {Jayasinghe}, {Ji}, {Johnson}, {Jones}, {J{\"o}nsson}, {Katkov}, {Khalatyan},
  {Kinemuchi}, {Kisku}, {Knapen}, {Kneib}, {Kollmeier}, {Kong}, {Kounkel},
  {Kreckel}, {Krishnarao}, {Lacerna}, {Lane}, {Langgin}, {Lavender}, {Law},
  {Lazarz}, {Leung}, {Leung}, {Lewis}, {Li}, {Li}, {Lian}, {Liang}, {Lin},
  {Lin}, {Lin}, {Lintott}, {Long}, {Longa-Pe{\~n}a}, {L{\'o}pez-Cob{\'a}},
  {Lu}, {Lundgren}, {Luo}, {Mackereth}, {de la Macorra}, {Mahadevan},
  {Majewski}, {Manchado}, {Mandeville}, {Maraston}, {Margalef-Bentabol},
  {Masseron}, {Masters}, {Mathur}, {McDermid}, {Mckay}, {Merloni},
  {Merrifield}, {Meszaros}, {Miglio}, {Di Mille}, {Minniti}, {Minsley},
  {Monachesi}, {Moon}, {Mosser}, {Mulchaey}, {Muna}, {Mu{\~n}oz}, {Myers},
  {Myers}, {Nadathur}, {Nair}, {Nandra}, {Neumann}, {Newman}, {Nidever},
  {Nikakhtar}, {Nitschelm}, {O'Connell}, {Garma-Oehmichen}, {Luan Souza de
  Oliveira}, {Olney}, {Oravetz}, {Ortigoza-Urdaneta}, {Osorio}, {Otter},
  {Pace}, {Padilla}, {Pan}, {Pan}, {Parikh}, {Parker}, {Peirani}, {Pe{\~n}a
  Ram{\'\i}rez}, {Penny}, {Percival}, {Perez-Fournon}, {Pinsonneault},
  {Poidevin}, {Poovelil}, {Price-Whelan}, {B{\'a}rbara de Andrade Queiroz},
  {Raddick}, {Ray}, {Rembold}, {Riddle}, {Riffel}, {Riffel}, {Rix}, {Robin},
  {Rodr{\'\i}guez-Puebla}, {Roman-Lopes}, {Rom{\'a}n-Z{\'u}{\~n}iga}, {Rose},
  {Ross}, {Rossi}, {Rubin}, {Salvato}, {S{\'a}nchez}, {S{\'a}nchez-Gallego},
  {Sanderson}, {Santana Rojas}, {Sarceno}, {Sarmiento}, {Sayres}, {Sazonova},
  {Schaefer}, {Schiavon}, {Schlegel}, {Schneider}, {Schultheis}, {Schwope},
  {Serenelli}, {Serna}, {Shao}, {Shapiro}, {Sharma}, {Shen}, {Shetrone}, {Shu},
  {Simon}, {Skrutskie}, {Smethurst}, {Smith}, {Sobeck}, {Spoo}, {Sprague},
  {Stark}, {Stassun}, {Steinmetz}, {Stello}, {Stone-Martinez},
  {Storchi-Bergmann}, {Stringfellow}, {Stutz}, {Su}, {Taghizadeh-Popp},
  {Talbot}, {Tayar}, {Telles}, {Teske}, {Thakar}, {Theissen}, {Tkachenko},
  {Thomas}, {Tojeiro}, {Hernandez Toledo}, {Troup}, {Trump}, {Trussler},
  {Turner}, {Tuttle}, {Unda-Sanzana}, {V{\'a}zquez-Mata}, {Valentini},
  {Valenzuela}, {Vargas-Gonz{\'a}lez}, {Vargas-Maga{\~n}a}, {Alfaro},
  {Villanova}, {Vincenzo}, {Wake}, {Warfield}, {Washington}, {Weaver},
  {Weijmans}, {Weinberg}, {Weiss}, {Westfall}, {Wild}, {Wilde}, {Wilson},
  {Wilson}, {Wilson}, {Wolf}, {Wood-Vasey}, {Yan}, {Zamora}, {Zasowski},
  {Zhang}, {Zhao}, {Zheng}, {Zheng}, \& {Zhu}}]{Abdurro22}
{Abdurro'uf}, {Accetta}, K., {Aerts}, C., {et~al.} 2022, \apjs, 259, 35,
  \dodoi{10.3847/1538-4365/ac4414}

\bibitem[{{Adams} {et~al.}(2015){Adams}, {Martini}, {Croxall}, {Overzier}, \&
  {Silverman}}]{Adams15}
{Adams}, S.~M., {Martini}, P., {Croxall}, K.~V., {Overzier}, R.~A., \&
  {Silverman}, J.~D. 2015, \mnras, 448, 1335, \dodoi{10.1093/mnras/stv065}

\bibitem[{{Alberts} \& {Noble}(2022)}]{Alberts22}
{Alberts}, S., \& {Noble}, A. 2022, Universe, 8, 554,
  \dodoi{10.3390/universe8110554}

\bibitem[{{An} {et~al.}(2013){An}, {Zheng}, {Meng}, {Chen}, {Wen}, \&
  {L{\"u}}}]{An13}
{An}, F., {Zheng}, X., {Meng}, Y., {et~al.} 2013, Science China Physics,
  Mechanics, and Astronomy, 56, 2226, \dodoi{10.1007/s11433-013-5331-y}

\bibitem[{{Ando} {et~al.}(2020){Ando}, {Shimasaku}, \& {Momose}}]{Ando20}
{Ando}, M., {Shimasaku}, K., \& {Momose}, R. 2020, \mnras, 496, 3169,
  \dodoi{10.1093/mnras/staa1757}

\bibitem[{{Angthopo} {et~al.}(2019){Angthopo}, {Ferreras}, \&
  {Silk}}]{Angthopo19}
{Angthopo}, J., {Ferreras}, I., \& {Silk}, J. 2019, \mnras, 488, L99,
  \dodoi{10.1093/mnrasl/slz106}

\bibitem[{{Annunziatella} {et~al.}(2018){Annunziatella}, {Marchesini},
  {Stefanon}, {Muzzin}, {Lange-Vagle}, {Cybulski}, {Labbe}, {Kado-Fong},
  {Bezanson}, {Brammer}, {Herrera}, {Lundgren}, {Marsan}, {Nonino}, {Rudnick},
  {Saracco}, {Tomer}, {Valdes}, {van der Burg}, {van Dokkum}, {Wake}, \&
  {Whitaker}}]{Annunziatella18}
{Annunziatella}, M., {Marchesini}, D., {Stefanon}, M., {et~al.} 2018, \pasp,
  130, 124501, \dodoi{10.1088/1538-3873/aae796}

\bibitem[{{Backhaus} {et~al.}(2024){Backhaus}, {Trump}, {Pirzkal}, {Barro},
  {Finkelstein}, {Arrabal Haro}, {Simons}, {Wessner}, {Cleri}, {Bagley},
  {Hirschmann}, {Nicholls}, {Dickinson}, {Kartaltepe}, {Papovich}, {Kocevski},
  {Koekemoer}, {Bisigello}, {Jaskot}, {Lucas}, {Jung}, {Wilkins}, {Yung},
  {Ferguson}, {Fontana}, {Grazian}, {Grogin}, {Kewley}, {Kirkpatrick}, {Lotz},
  {Pentericci}, {P{\'e}rez-Gonz{\'a}lez}, {Ravindranath}, {Somerville}, {Yang},
  {Holwerda}, {Kurczynski}, {Hathi}, {Rose}, \& {Davis}}]{Backhaus24}
{Backhaus}, B.~E., {Trump}, J.~R., {Pirzkal}, N., {et~al.} 2024, \apj, 962,
  195, \dodoi{10.3847/1538-4357/ad1520}

\bibitem[{{Bertin} \& {Arnouts}(1996)}]{Bertin96}
{Bertin}, E., \& {Arnouts}, S. 1996, \aaps, 117, 393,
  \dodoi{10.1051/aas:1996164}

\bibitem[{{Bielby} {et~al.}(2012){Bielby}, {Hudelot}, {McCracken}, {Ilbert},
  {Daddi}, {Le F{\`e}vre}, {Gonzalez-Perez}, {Kneib}, {Marmo}, {Mellier},
  {Salvato}, {Sanders}, \& {Willott}}]{Bielby12}
{Bielby}, R., {Hudelot}, P., {McCracken}, H.~J., {et~al.} 2012, \aap, 545, A23,
  \dodoi{10.1051/0004-6361/201118547}

\bibitem[{{Boquien} {et~al.}(2019){Boquien}, {Burgarella}, {Roehlly}, {Buat},
  {Ciesla}, {Corre}, {Inoue}, \& {Salas}}]{Boquien19}
{Boquien}, M., {Burgarella}, D., {Roehlly}, Y., {et~al.} 2019, \aap, 622, A103,
  \dodoi{10.1051/0004-6361/201834156}

\bibitem[{{Brammer} {et~al.}(2011){Brammer}, {Whitaker}, {van Dokkum},
  {Marchesini}, {Franx}, {Kriek}, {Labb{\'e}}, {Lee}, {Muzzin}, {Quadri},
  {Rudnick}, \& {Williams}}]{Brammer11}
{Brammer}, G.~B., {Whitaker}, K.~E., {van Dokkum}, P.~G., {et~al.} 2011, \apj,
  739, 24, \dodoi{10.1088/0004-637X/739/1/24}

\bibitem[{{Bruzual} \& {Charlot}(2003)}]{BC03}
{Bruzual}, G., \& {Charlot}, S. 2003, \mnras, 344, 1000,
  \dodoi{10.1046/j.1365-8711.2003.06897.x}

\bibitem[{{Bunker} {et~al.}(1995){Bunker}, {Warren}, {Hewett}, \&
  {Clements}}]{Bunker95}
{Bunker}, A.~J., {Warren}, S.~J., {Hewett}, P.~C., \& {Clements}, D.~L. 1995,
  \mnras, 273, 513, \dodoi{10.1093/mnras/273.2.513}

\bibitem[{{Calzetti} {et~al.}(2000){Calzetti}, {Armus}, {Bohlin}, {Kinney},
  {Koornneef}, \& {Storchi-Bergmann}}]{Calzetti00}
{Calzetti}, D., {Armus}, L., {Bohlin}, R.~C., {et~al.} 2000, \apj, 533, 682,
  \dodoi{10.1086/308692}

\bibitem[{{Chabrier}(2003)}]{Chabrier03}
{Chabrier}, G. 2003, \pasp, 115, 763, \dodoi{10.1086/376392}

\bibitem[{{Chiang} {et~al.}(2013){Chiang}, {Overzier}, \&
  {Gebhardt}}]{Chiang13}
{Chiang}, Y.-K., {Overzier}, R., \& {Gebhardt}, K. 2013, \apj, 779, 127,
  \dodoi{10.1088/0004-637X/779/2/127}

\bibitem[{{Chiang} {et~al.}(2017){Chiang}, {Overzier}, {Gebhardt}, \&
  {Henriques}}]{Chiang17}
{Chiang}, Y.-K., {Overzier}, R.~A., {Gebhardt}, K., \& {Henriques}, B. 2017,
  \apjl, 844, L23, \dodoi{10.3847/2041-8213/aa7e7b}

\bibitem[{{Cooke} {et~al.}(2014){Cooke}, {Hatch}, {Muldrew}, {Rigby}, \&
  {Kurk}}]{Cooke14}
{Cooke}, E.~A., {Hatch}, N.~A., {Muldrew}, S.~I., {Rigby}, E.~E., \& {Kurk},
  J.~D. 2014, \mnras, 440, 3262, \dodoi{10.1093/mnras/stu522}

\bibitem[{{Cucciati} {et~al.}(2014){Cucciati}, {Zamorani}, {Lemaux},
  {Bardelli}, {Cimatti}, {Le F{\`e}vre}, {Cassata}, {Garilli}, {Le Brun},
  {Maccagni}, {Pentericci}, {Tasca}, {Thomas}, {Vanzella}, {Zucca}, {Amorin},
  {Capak}, {Cassar{\`a}}, {Castellano}, {Cuby}, {de la Torre}, {Durkalec},
  {Fontana}, {Giavalisco}, {Grazian}, {Hathi}, {Ilbert}, {Moreau}, {Paltani},
  {Ribeiro}, {Salvato}, {Schaerer}, {Scodeggio}, {Sommariva}, {Talia},
  {Taniguchi}, {Tresse}, {Vergani}, {Wang}, {Charlot}, {Contini}, {Fotopoulou},
  {L{\'o}pez-Sanjuan}, {Mellier}, \& {Scoville}}]{Cucciati14}
{Cucciati}, O., {Zamorani}, G., {Lemaux}, B.~C., {et~al.} 2014, \aap, 570, A16,
  \dodoi{10.1051/0004-6361/201423811}

\bibitem[{{Cucciati} {et~al.}(2018){Cucciati}, {Lemaux}, {Zamorani}, {Le
  F{\`e}vre}, {Tasca}, {Hathi}, {Lee}, {Bardelli}, {Cassata}, {Garilli}, {Le
  Brun}, {Maccagni}, {Pentericci}, {Thomas}, {Vanzella}, {Zucca}, {Lubin},
  {Amorin}, {Cassar{\`a}}, {Cimatti}, {Talia}, {Vergani}, {Koekemoer}, {Pforr},
  \& {Salvato}}]{Cucciati18}
{Cucciati}, O., {Lemaux}, B.~C., {Zamorani}, G., {et~al.} 2018, \aap, 619, A49,
  \dodoi{10.1051/0004-6361/201833655}

\bibitem[{{Daddi} {et~al.}(2007){Daddi}, {Dickinson}, {Morrison}, {Chary},
  {Cimatti}, {Elbaz}, {Frayer}, {Renzini}, {Pope}, {Alexander}, {Bauer},
  {Giavalisco}, {Huynh}, {Kurk}, \& {Mignoli}}]{Daddi07}
{Daddi}, E., {Dickinson}, M., {Morrison}, G., {et~al.} 2007, \apj, 670, 156,
  \dodoi{10.1086/521818}

\bibitem[{{Daikuhara} {et~al.}(2024){Daikuhara}, {Kodama},
  {P{\'e}rez-Mart{\'\i}nez}, {Shimakawa}, {Suzuki}, {Tadaki}, {Koyama}, \&
  {Tanaka}}]{Daikuhara24}
{Daikuhara}, K., {Kodama}, T., {P{\'e}rez-Mart{\'\i}nez}, J.~M., {et~al.} 2024,
  \mnras, 531, 2335, \dodoi{10.1093/mnras/stae1243}

\bibitem[{{Dale} {et~al.}(2014){Dale}, {Helou}, {Magdis}, {Armus},
  {D{\'\i}az-Santos}, \& {Shi}}]{Dale14}
{Dale}, D.~A., {Helou}, G., {Magdis}, G.~E., {et~al.} 2014, \apj, 784, 83,
  \dodoi{10.1088/0004-637X/784/1/83}

\bibitem[{{Dey} {et~al.}(2016){Dey}, {Lee}, {Reddy}, {Cooper}, {Inami}, {Hong},
  {Gonzalez}, \& {Jannuzi}}]{Dey16}
{Dey}, A., {Lee}, K.-S., {Reddy}, N., {et~al.} 2016, \apj, 823, 11,
  \dodoi{10.3847/0004-637X/823/1/11}

\bibitem[{{Dressler}(1980)}]{Dressler80}
{Dressler}, A. 1980, \apj, 236, 351, \dodoi{10.1086/157753}

\bibitem[{{Dressler} {et~al.}(1997){Dressler}, {Oemler}, {Couch}, {Smail},
  {Ellis}, {Barger}, {Butcher}, {Poggianti}, \& {Sharples}}]{Dressler97}
{Dressler}, A., {Oemler}, Augustus, J., {Couch}, W.~J., {et~al.} 1997, \apj,
  490, 577, \dodoi{10.1086/304890}

\bibitem[{{Elbaz} {et~al.}(2007){Elbaz}, {Daddi}, {Le Borgne}, {Dickinson},
  {Alexander}, {Chary}, {Starck}, {Brand t}, {Kitzbichler}, {MacDonald},
  {Nonino}, {Popesso}, {Stern}, \& {Vanzella}}]{Elbaz07}
{Elbaz}, D., {Daddi}, E., {Le Borgne}, D., {et~al.} 2007, \aap, 468, 33,
  \dodoi{10.1051/0004-6361:20077525}

\bibitem[{{Faisst} {et~al.}(2016){Faisst}, {Capak}, {Hsieh}, {Laigle},
  {Salvato}, {Tasca}, {Cassata}, {Davidzon}, {Ilbert}, {Le F{\`e}vre},
  {Masters}, {McCracken}, {Steinhardt}, {Silverman}, {de Barros}, {Hasinger},
  \& {Scoville}}]{Faisst16}
{Faisst}, A.~L., {Capak}, P., {Hsieh}, B.~C., {et~al.} 2016, \apj, 821, 122,
  \dodoi{10.3847/0004-637X/821/2/122}

\bibitem[{{Finkelstein} {et~al.}(2009){Finkelstein}, {Rhoads}, {Malhotra}, \&
  {Grogin}}]{Finkelstein09}
{Finkelstein}, S.~L., {Rhoads}, J.~E., {Malhotra}, S., \& {Grogin}, N. 2009,
  \apj, 691, 465, \dodoi{10.1088/0004-637X/691/1/465}

\bibitem[{{Forrest} {et~al.}(2017){Forrest}, {Tran}, {Broussard}, {Allen},
  {Apfel}, {Cowley}, {Glazebrook}, {Kacprzak}, {Labb{\'e}}, {Nanayakkara},
  {Papovich}, {Quadri}, {Spitler}, {Straatman}, \& {Tomczak}}]{Forrest17}
{Forrest}, B., {Tran}, K.-V.~H., {Broussard}, A., {et~al.} 2017, \apjl, 838,
  L12, \dodoi{10.3847/2041-8213/aa653b}

\bibitem[{{Gao} {et~al.}(2005){Gao}, {Springel}, \& {White}}]{Gao05}
{Gao}, L., {Springel}, V., \& {White}, S. D.~M. 2005, \mnras, 363, L66,
  \dodoi{10.1111/j.1745-3933.2005.00084.x}

\bibitem[{{Goto} {et~al.}(2003){Goto}, {Yamauchi}, {Fujita}, {Okamura},
  {Sekiguchi}, {Smail}, {Bernardi}, \& {Gomez}}]{Goto03}
{Goto}, T., {Yamauchi}, C., {Fujita}, Y., {et~al.} 2003, \mnras, 346, 601,
  \dodoi{10.1046/j.1365-2966.2003.07114.x}

\bibitem[{{Gwyn}(2012)}]{Gwyn12}
{Gwyn}, S. D.~J. 2012, \aj, 143, 38, \dodoi{10.1088/0004-6256/143/2/38}

\bibitem[{{Harikane} {et~al.}(2019){Harikane}, {Ouchi}, {Ono}, {Fujimoto},
  {Donevski}, {Shibuya}, {Faisst}, {Goto}, {Hatsukade}, {Kashikawa}, {Kohno},
  {Hashimoto}, {Higuchi}, {Inoue}, {Lin}, {Martin}, {Overzier}, {Smail},
  {Toshikawa}, {Umehata}, {Ao}, {Chapman}, {Clements}, {Im}, {Jing},
  {Kawaguchi}, {Lee}, {Lee}, {Lin}, {Matsuoka}, {Marinello}, {Nagao},
  {Onodera}, {Toft}, \& {Wang}}]{Harikane19}
{Harikane}, Y., {Ouchi}, M., {Ono}, Y., {et~al.} 2019, \apj, 883, 142,
  \dodoi{10.3847/1538-4357/ab2cd5}

\bibitem[{{Hatch} {et~al.}(2011){Hatch}, {De Breuck}, {Galametz}, {Miley},
  {Overzier}, {R{\"o}ttgering}, {Doherty}, {Kodama}, {Kurk}, {Seymour},
  {Venemans}, {Vernet}, \& {Zirm}}]{Hatch11}
{Hatch}, N.~A., {De Breuck}, C., {Galametz}, A., {et~al.} 2011, \mnras, 410,
  1537, \dodoi{10.1111/j.1365-2966.2010.17538.x}

\bibitem[{{Hayashi} {et~al.}(2012){Hayashi}, {Kodama}, {Tadaki}, {Koyama}, \&
  {Tanaka}}]{Hayashi12}
{Hayashi}, M., {Kodama}, T., {Tadaki}, K.-i., {Koyama}, Y., \& {Tanaka}, I.
  2012, \apj, 757, 15, \dodoi{10.1088/0004-637X/757/1/15}

\bibitem[{{Hayashi} {et~al.}(2016){Hayashi}, {Kodama}, {Tanaka}, {Shimakawa},
  {Koyama}, {Tadaki}, {Suzuki}, \& {Yamamoto}}]{Hayashi16}
{Hayashi}, M., {Kodama}, T., {Tanaka}, I., {et~al.} 2016, \apjl, 826, L28,
  \dodoi{10.3847/2041-8205/826/2/L28}

\bibitem[{{Henriques} {et~al.}(2012){Henriques}, {White}, {Lemson}, {Thomas},
  {Guo}, {Marleau}, \& {Overzier}}]{Henriques12}
{Henriques}, B. M.~B., {White}, S. D.~M., {Lemson}, G., {et~al.} 2012, \mnras,
  421, 2904, \dodoi{10.1111/j.1365-2966.2012.20521.x}

\bibitem[{{Hung} {et~al.}(2025){Hung}, {Lemaux}, {Cucciati}, {Forrest}, {Shah},
  {Gal}, {Giddings}, {Sikorski}, {Golden-Marx}, {Lubin}, {Hathi}, {Zamorani},
  {Shen}, {Bardelli}, {Cassar{\`a}}, {De Lucia}, {Fontanot}, {Garilli},
  {Guaita}, {Hirschmann}, {Lee}, {Newman}, {Ramakrishnan}, {Vergani}, {Xie}, \&
  {Zucca}}]{Hung25}
{Hung}, D., {Lemaux}, B.~C., {Cucciati}, O., {et~al.} 2025, \apj, 980, 155,
  \dodoi{10.3847/1538-4357/ada616}

\bibitem[{{Husband} {et~al.}(2013){Husband}, {Bremer}, {Stanway}, {Davies},
  {Lehnert}, \& {Douglas}}]{Husband13}
{Husband}, K., {Bremer}, M.~N., {Stanway}, E.~R., {et~al.} 2013, \mnras, 432,
  2869, \dodoi{10.1093/mnras/stt642}

\bibitem[{{Ito} {et~al.}(2020){Ito}, {Kashikawa}, {Toshikawa}, {Overzier},
  {Kubo}, {Uchiyama}, {Liang}, {Onoue}, {Tanaka}, {Komiyama}, {Lee}, {Lin},
  {Marinello}, {Martin}, \& {Shibuya}}]{Ito20}
{Ito}, K., {Kashikawa}, N., {Toshikawa}, J., {et~al.} 2020, \apj, 899, 5,
  \dodoi{10.3847/1538-4357/aba269}

\bibitem[{{Ito} {et~al.}(2021){Ito}, {Kashikawa}, {Tanaka}, {Kubo}, {Liang},
  {Toshikawa}, {Uchiyama}, {Ishimoto}, {Yoshioka}, \& {Takeda}}]{Ito21}
{Ito}, K., {Kashikawa}, N., {Tanaka}, M., {et~al.} 2021, \apj, 916, 35,
  \dodoi{10.3847/1538-4357/abfc50}

\bibitem[{{Ito} {et~al.}(2023){Ito}, {Tanaka}, {Valentino}, {Toft}, {Brammer},
  {Gould}, {Ilbert}, {Kashikawa}, {Kubo}, {Liang}, {McCracken}, \&
  {Weaver}}]{Ito23}
{Ito}, K., {Tanaka}, M., {Valentino}, F., {et~al.} 2023, \apjl, 945, L9,
  \dodoi{10.3847/2041-8213/acb49b}

\bibitem[{{Jiang} {et~al.}(2018){Jiang}, {Wu}, {Bian}, {Chiang}, {Ho}, {Shen},
  {Zheng}, {Bailey}, {Blanc}, {Crane}, {Fan}, {Mateo}, {Olszewski},
  {Oyarz{\'u}n}, {Wang}, \& {Wu}}]{Jiang18}
{Jiang}, L., {Wu}, J., {Bian}, F., {et~al.} 2018, Nature Astronomy, 2, 962,
  \dodoi{10.1038/s41550-018-0587-9}

\bibitem[{{Jin} {et~al.}(2024){Jin}, {Sillassen}, {Magdis}, {Brinch},
  {Shuntov}, {Brammer}, {Gobat}, {Valentino}, {Carnall}, {Lee}, {Vijayan},
  {Gillman}, {Kokorev}, {Le Bail}, {Greve}, {Gullberg}, {Gould}, \&
  {Toft}}]{Jin24}
{Jin}, S., {Sillassen}, N.~B., {Magdis}, G.~E., {et~al.} 2024, \aap, 683, L4,
  \dodoi{10.1051/0004-6361/202348540}

\bibitem[{{Kashikawa} {et~al.}(2007){Kashikawa}, {Kitayama}, {Doi}, {Misawa},
  {Komiyama}, \& {Ota}}]{Kashikawa07}
{Kashikawa}, N., {Kitayama}, T., {Doi}, M., {et~al.} 2007, \apj, 663, 765,
  \dodoi{10.1086/518410}

\bibitem[{{Kauffmann} {et~al.}(2004){Kauffmann}, {White}, {Heckman},
  {M{\'e}nard}, {Brinchmann}, {Charlot}, {Tremonti}, \&
  {Brinkmann}}]{Kauffmann04}
{Kauffmann}, G., {White}, S. D.~M., {Heckman}, T.~M., {et~al.} 2004, \mnras,
  353, 713, \dodoi{10.1111/j.1365-2966.2004.08117.x}

\bibitem[{{Khostovan} {et~al.}(2015){Khostovan}, {Sobral}, {Mobasher}, {Best},
  {Smail}, {Stott}, {Hemmati}, \& {Nayyeri}}]{Khostovan15}
{Khostovan}, A.~A., {Sobral}, D., {Mobasher}, B., {et~al.} 2015, \mnras, 452,
  3948, \dodoi{10.1093/mnras/stv1474}

\bibitem[{{Koyama} {et~al.}(2013){Koyama}, {Smail}, {Kurk}, {Geach}, {Sobral},
  {Kodama}, {Nakata}, {Swinbank}, {Best}, {Hayashi}, \& {Tadaki}}]{Koyama13}
{Koyama}, Y., {Smail}, I., {Kurk}, J., {et~al.} 2013, \mnras, 434, 423,
  \dodoi{10.1093/mnras/stt1035}

\bibitem[{{Koyama} {et~al.}(2021){Koyama}, {Polletta}, {Tanaka}, {Kodama},
  {Dole}, {Soucail}, {Frye}, {Lehnert}, \& {Scodeggio}}]{Koyama20}
{Koyama}, Y., {Polletta}, M. d.~C., {Tanaka}, I., {et~al.} 2021, \mnras, 503,
  L1, \dodoi{10.1093/mnrasl/slab013}

\bibitem[{{Kubo} {et~al.}(2013){Kubo}, {Uchimoto}, {Yamada}, {Kajisawa},
  {Ichikawa}, {Matsuda}, {Akiyama}, {Hayashino}, {Konishi}, {Nishimura},
  {Omata}, {Suzuki}, {Tanaka}, {Yoshikawa}, {Alexander}, {Fazio}, {Huang}, \&
  {Lehmer}}]{Kubo13}
{Kubo}, M., {Uchimoto}, Y.~K., {Yamada}, T., {et~al.} 2013, \apj, 778, 170,
  \dodoi{10.1088/0004-637X/778/2/170}

\bibitem[{{Kuiper} {et~al.}(2011){Kuiper}, {Hatch}, {Venemans}, {Miley},
  {R{\"o}ttgering}, {Kurk}, {Overzier}, {Pentericci}, {Bland-Hawthorn}, \&
  {Cepa}}]{Kuiper11}
{Kuiper}, E., {Hatch}, N.~A., {Venemans}, B.~P., {et~al.} 2011, \mnras, 417,
  1088, \dodoi{10.1111/j.1365-2966.2011.19324.x}

\bibitem[{{Kurk} {et~al.}(2004){Kurk}, {Pentericci}, {R{\"o}ttgering}, \&
  {Miley}}]{Kurk04}
{Kurk}, J.~D., {Pentericci}, L., {R{\"o}ttgering}, H.~J.~A., \& {Miley}, G.~K.
  2004, \aap, 428, 793, \dodoi{10.1051/0004-6361:20040075}

\bibitem[{{Labb{\'e}} {et~al.}(2005){Labb{\'e}}, {Huang}, {Franx}, {Rudnick},
  {Barmby}, {Daddi}, {van Dokkum}, {Fazio}, {Schreiber}, {Moorwood}, {Rix},
  {R{\"o}ttgering}, {Trujillo}, \& {van der Werf}}]{Labbe05}
{Labb{\'e}}, I., {Huang}, J., {Franx}, M., {et~al.} 2005, \apjl, 624, L81,
  \dodoi{10.1086/430700}

\bibitem[{{Landy} \& {Szalay}(1993)}]{Landy93}
{Landy}, S.~D., \& {Szalay}, A.~S. 1993, \apj, 412, 64, \dodoi{10.1086/172900}

\bibitem[{{Lee} {et~al.}(2014){Lee}, {Dey}, {Hong}, {Reddy}, {Wilson},
  {Jannuzi}, {Inami}, \& {Gonzalez}}]{Lee14}
{Lee}, K.-S., {Dey}, A., {Hong}, S., {et~al.} 2014, \apj, 796, 126,
  \dodoi{10.1088/0004-637X/796/2/126}

\bibitem[{{Lemaux} {et~al.}(2014){Lemaux}, {Cucciati}, {Tasca}, {Le F{\`e}vre},
  {Zamorani}, {Cassata}, {Garilli}, {Le Brun}, {Maccagni}, {Pentericci},
  {Thomas}, {Vanzella}, {Zucca}, {Amor{\'\i}n}, {Bardelli}, {Capak},
  {Cassar{\`a}}, {Castellano}, {Cimatti}, {Cuby}, {de la Torre}, {Durkalec},
  {Fontana}, {Giavalisco}, {Grazian}, {Hathi}, {Ilbert}, {Moreau}, {Paltani},
  {Ribeiro}, {Salvato}, {Schaerer}, {Scodeggio}, {Sommariva}, {Talia},
  {Taniguchi}, {Tresse}, {Vergani}, {Wang}, {Charlot}, {Contini}, {Fotopoulou},
  {Gal}, {Kocevski}, {L{\'o}pez-Sanjuan}, {Lubin}, {Mellier}, {Sadibekova}, \&
  {Scoville}}]{Lemaux14}
{Lemaux}, B.~C., {Cucciati}, O., {Tasca}, L.~A.~M., {et~al.} 2014, \aap, 572,
  A41, \dodoi{10.1051/0004-6361/201423828}

\bibitem[{{Lemaux} {et~al.}(2018){Lemaux}, {Le F{\`e}vre}, {Cucciati},
  {Ribeiro}, {Tasca}, {Zamorani}, {Ilbert}, {Thomas}, {Bardelli}, {Cassata},
  {Hathi}, {Pforr}, {Smol{\v{c}}i{\'c}}, {Delvecchio}, {Novak}, {Berta},
  {McCracken}, {Koekemoer}, {Amor{\'\i}n}, {Garilli}, {Maccagni}, {Schaerer},
  \& {Zucca}}]{Lemaux18}
{Lemaux}, B.~C., {Le F{\`e}vre}, O., {Cucciati}, O., {et~al.} 2018, \aap, 615,
  A77, \dodoi{10.1051/0004-6361/201730870}

\bibitem[{{Lemaux} {et~al.}(2022){Lemaux}, {Cucciati}, {Le F{\`e}vre},
  {Zamorani}, {Lubin}, {Hathi}, {Ilbert}, {Pelliccia}, {Amor{\'\i}n},
  {Bardelli}, {Cassata}, {Gal}, {Garilli}, {Guaita}, {Giavalisco}, {Hung},
  {Koekemoer}, {Maccagni}, {Pentericci}, {Ribeiro}, {Schaerer}, {Shah}, {Shen},
  {Staab}, {Talia}, {Thomas}, {Tomczak}, {Tresse}, {Vanzella}, {Vergani}, \&
  {Zucca}}]{Lemaux22}
{Lemaux}, B.~C., {Cucciati}, O., {Le F{\`e}vre}, O., {et~al.} 2022, \aap, 662,
  A33, \dodoi{10.1051/0004-6361/202039346}

\bibitem[{{Li} {et~al.}(2008){Li}, {Mo}, \& {Gao}}]{Li08}
{Li}, Y., {Mo}, H.~J., \& {Gao}, L. 2008, \mnras, 389, 1419,
  \dodoi{10.1111/j.1365-2966.2008.13667.x}

\bibitem[{{Mart{\'\i}n-Navarro} {et~al.}(2018){Mart{\'\i}n-Navarro},
  {Vazdekis}, {Falc{\'o}n-Barroso}, {La Barbera}, {Y{\i}ld{\i}r{\i}m}, \& {van
  de Ven}}]{Mart18}
{Mart{\'\i}n-Navarro}, I., {Vazdekis}, A., {Falc{\'o}n-Barroso}, J., {et~al.}
  2018, \mnras, 475, 3700, \dodoi{10.1093/mnras/stx3346}

\bibitem[{{Maschietto} {et~al.}(2008){Maschietto}, {Hatch}, {Venemans},
  {R{\"o}ttgering}, {Miley}, {Overzier}, {Dopita}, {Eisenhardt}, {Kurk},
  {Meurer}, {Pentericci}, {Rosati}, {Stanford}, {van Breugel}, \&
  {Zirm}}]{Maschietto08}
{Maschietto}, F., {Hatch}, N.~A., {Venemans}, B.~P., {et~al.} 2008, \mnras,
  389, 1223, \dodoi{10.1111/j.1365-2966.2008.13571.x}

\bibitem[{{McConachie} {et~al.}(2022){McConachie}, {Wilson}, {Forrest},
  {Marsan}, {Muzzin}, {Cooper}, {Annunziatella}, {Marchesini}, {Chan}, {Gomez},
  {Abdullah}, {Saracco}, \& {Nantais}}]{McConachie22}
{McConachie}, I., {Wilson}, G., {Forrest}, B., {et~al.} 2022, \apj, 926, 37,
  \dodoi{10.3847/1538-4357/ac2b9f}

\bibitem[{{Mei} {et~al.}(2023){Mei}, {Hatch}, {Amodeo}, {Afanasiev}, {De
  Breuck}, {Stern}, {Cooke}, {Gonzalez}, {Noirot}, {Rettura}, {Seymour},
  {Stanford}, {Vernet}, \& {Wylezalek}}]{Mei23}
{Mei}, S., {Hatch}, N.~A., {Amodeo}, S., {et~al.} 2023, \aap, 670, A58,
  \dodoi{10.1051/0004-6361/202243551}

\bibitem[{{Merlin} {et~al.}(2015){Merlin}, {Fontana}, {Ferguson}, {Dunlop},
  {Elbaz}, {Bourne}, {Bruce}, {Buitrago}, {Castellano}, {Schreiber}, {Grazian},
  {McLure}, {Okumura}, {Shu}, {Wang}, {Amor{\'\i}n}, {Boutsia}, {Cappelluti},
  {Comastri}, {Derriere}, {Faber}, \& {Santini}}]{Merlin15}
{Merlin}, E., {Fontana}, A., {Ferguson}, H.~C., {et~al.} 2015, \aap, 582, A15,
  \dodoi{10.1051/0004-6361/201526471}

\bibitem[{{Merlin} {et~al.}(2016){Merlin}, {Bourne}, {Castellano}, {Ferguson},
  {Wang}, {Derriere}, {Dunlop}, {Elbaz}, \& {Fontana}}]{Merlin16}
{Merlin}, E., {Bourne}, N., {Castellano}, M., {et~al.} 2016, \aap, 595, A97,
  \dodoi{10.1051/0004-6361/201628751}

\bibitem[{{Muzzin} {et~al.}(2013){Muzzin}, {Marchesini}, {Stefanon}, {Franx},
  {McCracken}, {Milvang-Jensen}, {Dunlop}, {Fynbo}, {Brammer}, {Labb{\'e}}, \&
  {van Dokkum}}]{Muzzin13}
{Muzzin}, A., {Marchesini}, D., {Stefanon}, M., {et~al.} 2013, \apj, 777, 18,
  \dodoi{10.1088/0004-637X/777/1/18}

\bibitem[{{Naufal} {et~al.}(2024){Naufal}, {Koyama}, {D'Eugenio},
  {Dannerbauer}, {Shimakawa}, {P{\'e}rez-Mart{\'\i}nez}, {Kodama}, {Zhang}, \&
  {Daikuhara}}]{Naufal24}
{Naufal}, A., {Koyama}, Y., {D'Eugenio}, C., {et~al.} 2024, \apj, 977, 58,
  \dodoi{10.3847/1538-4357/ad8dcf}

\bibitem[{{Noeske} {et~al.}(2007){Noeske}, {Weiner}, {Faber}, {Papovich},
  {Koo}, {Somerville}, {Bundy}, {Conselice}, {Newman}, {Schiminovich}, {Le
  Floc'h}, {Coil}, {Rieke}, {Lotz}, {Primack}, {Barmby}, {Cooper}, {Davis},
  {Ellis}, {Fazio}, {Guhathakurta}, {Huang}, {Kassin}, {Martin}, {Phillips},
  {Rich}, {Small}, {Willmer}, \& {Wilson}}]{Noeske07}
{Noeske}, K.~G., {Weiner}, B.~J., {Faber}, S.~M., {et~al.} 2007, \apjl, 660,
  L43, \dodoi{10.1086/517926}

\bibitem[{{Noll} {et~al.}(2009){Noll}, {Burgarella}, {Giovannoli}, {Buat},
  {Marcillac}, \& {Mu{\~n}oz-Mateos}}]{Noll09}
{Noll}, S., {Burgarella}, D., {Giovannoli}, E., {et~al.} 2009, \aap, 507, 1793,
  \dodoi{10.1051/0004-6361/200912497}

\bibitem[{{Oke} \& {Gunn}(1983)}]{Oke83}
{Oke}, J.~B., \& {Gunn}, J.~E. 1983, \apj, 266, 713, \dodoi{10.1086/160817}

\bibitem[{{Onodera} {et~al.}(2016){Onodera}, {Carollo}, {Lilly}, {Renzini},
  {Arimoto}, {Capak}, {Daddi}, {Scoville}, {Tacchella}, {Tatehora}, \&
  {Zamorani}}]{Onodera16}
{Onodera}, M., {Carollo}, C.~M., {Lilly}, S., {et~al.} 2016, \apj, 822, 42,
  \dodoi{10.3847/0004-637X/822/1/42}

\bibitem[{{Onodera} {et~al.}(2020){Onodera}, {Shimakawa}, {Suzuki}, {Tanaka},
  {Harikane}, {Hayashi}, {Kodama}, {Koyama}, {Nakajima}, \&
  {Shibuya}}]{Onodera20}
{Onodera}, M., {Shimakawa}, R., {Suzuki}, T.~L., {et~al.} 2020, \apj, 904, 180,
  \dodoi{10.3847/1538-4357/abc174}

\bibitem[{{Overzier}(2016)}]{Overzier16}
{Overzier}, R.~A. 2016, \aapr, 24, 14, \dodoi{10.1007/s00159-016-0100-3}

\bibitem[{{Pentericci} {et~al.}(2000){Pentericci}, {Kurk}, {R{\"o}ttgering},
  {Miley}, {van Breugel}, {Carilli}, {Ford}, {Heckman}, {McCarthy}, \&
  {Moorwood}}]{Pentericci00}
{Pentericci}, L., {Kurk}, J.~D., {R{\"o}ttgering}, H.~J.~A., {et~al.} 2000,
  \aap, 361, L25.
\newblock \doarXiv{astro-ph/0008143}

\bibitem[{{P{\'e}rez-Mart{\'\i}nez} {et~al.}(2023){P{\'e}rez-Mart{\'\i}nez},
  {Dannerbauer}, {Kodama}, {Koyama}, {Shimakawa}, {Suzuki}, {Calvi}, {Chen},
  {Daikuhara}, {Hatch}, {Laza-Ramos}, {Sobral}, {Stott}, \& {Tanaka}}]{Perez23}
{P{\'e}rez-Mart{\'\i}nez}, J.~M., {Dannerbauer}, H., {Kodama}, T., {et~al.}
  2023, \mnras, 518, 1707, \dodoi{10.1093/mnras/stac2784}

\bibitem[{{P{\'e}rez-Mart{\'\i}nez} {et~al.}(2024){P{\'e}rez-Mart{\'\i}nez},
  {Kodama}, {Koyama}, {Shimakawa}, {Suzuki}, {Daikuhara}, {Adachi}, {Onodera},
  \& {Tanaka}}]{Perez24}
{P{\'e}rez-Mart{\'\i}nez}, J.~M., {Kodama}, T., {Koyama}, Y., {et~al.} 2024,
  \mnras, 527, 10221, \dodoi{10.1093/mnras/stad3805}

\bibitem[{{Planck Collaboration} {et~al.}(2020){Planck Collaboration},
  {Aghanim}, {Akrami}, {Ashdown}, {Aumont}, {Baccigalupi}, {Ballardini},
  {Banday}, {Barreiro}, {Bartolo}, {Basak}, {Battye}, {Benabed}, {Bernard},
  {Bersanelli}, {Bielewicz}, {Bock}, {Bond}, {Borrill}, {Bouchet}, {Boulanger},
  {Bucher}, {Burigana}, {Butler}, {Calabrese}, {Cardoso}, {Carron},
  {Challinor}, {Chiang}, {Chluba}, {Colombo}, {Combet}, {Contreras}, {Crill},
  {Cuttaia}, {de Bernardis}, {de Zotti}, {Delabrouille}, {Delouis}, {Di
  Valentino}, {Diego}, {Dor{\'e}}, {Douspis}, {Ducout}, {Dupac}, {Dusini},
  {Efstathiou}, {Elsner}, {En{\ss}lin}, {Eriksen}, {Fantaye}, {Farhang},
  {Fergusson}, {Fernandez-Cobos}, {Finelli}, {Forastieri}, {Frailis},
  {Fraisse}, {Franceschi}, {Frolov}, {Galeotta}, {Galli}, {Ganga},
  {G{\'e}nova-Santos}, {Gerbino}, {Ghosh}, {Gonz{\'a}lez-Nuevo}, {G{\'o}rski},
  {Gratton}, {Gruppuso}, {Gudmundsson}, {Hamann}, {Handley}, {Hansen},
  {Herranz}, {Hildebrandt}, {Hivon}, {Huang}, {Jaffe}, {Jones}, {Karakci},
  {Keih{\"a}nen}, {Keskitalo}, {Kiiveri}, {Kim}, {Kisner}, {Knox},
  {Krachmalnicoff}, {Kunz}, {Kurki-Suonio}, {Lagache}, {Lamarre}, {Lasenby},
  {Lattanzi}, {Lawrence}, {Le Jeune}, {Lemos}, {Lesgourgues}, {Levrier},
  {Lewis}, {Liguori}, {Lilje}, {Lilley}, {Lindholm}, {L{\'o}pez-Caniego},
  {Lubin}, {Ma}, {Mac{\'\i}as-P{\'e}rez}, {Maggio}, {Maino}, {Mandolesi},
  {Mangilli}, {Marcos-Caballero}, {Maris}, {Martin}, {Martinelli},
  {Mart{\'\i}nez-Gonz{\'a}lez}, {Matarrese}, {Mauri}, {McEwen}, {Meinhold},
  {Melchiorri}, {Mennella}, {Migliaccio}, {Millea}, {Mitra},
  {Miville-Desch{\^e}nes}, {Molinari}, {Montier}, {Morgante}, {Moss}, {Natoli},
  {N{\o}rgaard-Nielsen}, {Pagano}, {Paoletti}, {Partridge}, {Patanchon},
  {Peiris}, {Perrotta}, {Pettorino}, {Piacentini}, {Polastri}, {Polenta},
  {Puget}, {Rachen}, {Reinecke}, {Remazeilles}, {Renzi}, {Rocha}, {Rosset},
  {Roudier}, {Rubi{\~n}o-Mart{\'\i}n}, {Ruiz-Granados}, {Salvati}, {Sandri},
  {Savelainen}, {Scott}, {Shellard}, {Sirignano}, {Sirri}, {Spencer},
  {Sunyaev}, {Suur-Uski}, {Tauber}, {Tavagnacco}, {Tenti}, {Toffolatti},
  {Tomasi}, {Trombetti}, {Valenziano}, {Valiviita}, {Van Tent}, {Vibert},
  {Vielva}, {Villa}, {Vittorio}, {Wandelt}, {Wehus}, {White}, {White},
  {Zacchei}, \& {Zonca}}]{Planck20}
{Planck Collaboration}, {Aghanim}, N., {Akrami}, Y., {et~al.} 2020, \aap, 641,
  A6, \dodoi{10.1051/0004-6361/201833910}

\bibitem[{{Polletta} {et~al.}(2021){Polletta}, {Soucail}, {Dole}, {Lehnert},
  {Pointecouteau}, {Vietri}, {Scodeggio}, {Montier}, {Koyama}, {Lagache},
  {Frye}, {Cusano}, \& {Fumana}}]{Polletta21}
{Polletta}, M., {Soucail}, G., {Dole}, H., {et~al.} 2021, \aap, 654, A121,
  \dodoi{10.1051/0004-6361/202140612}

\bibitem[{Richardson(1972)}]{Richardson:72}
Richardson, W.~H. 1972, J. Opt. Soc. Am., 62, 55,
  \dodoi{10.1364/JOSA.62.000055}

\bibitem[{{Salim}(2014)}]{Salim14}
{Salim}, S. 2014, Serbian Astronomical Journal, 189, 1,
  \dodoi{10.2298/SAJ1489001S}

\bibitem[{{Shi} {et~al.}(2024{\natexlab{a}}){Shi}, {Wang}, {Zheng}, {Cai},
  {Fan}, {Bian}, \& {Teplitz}}]{Shi24b}
{Shi}, D.~D., {Wang}, X., {Zheng}, X.~Z., {et~al.} 2024{\natexlab{a}}, \apj,
  963, 21, \dodoi{10.3847/1538-4357/ad17c3}

\bibitem[{{Shi} {et~al.}(2024{\natexlab{b}}){Shi}, {Malavasi}, {Toshikawa}, \&
  {Zheng}}]{Shi24a}
{Shi}, K., {Malavasi}, N., {Toshikawa}, J., \& {Zheng}, X. 2024{\natexlab{b}},
  \apj, 961, 39, \dodoi{10.3847/1538-4357/ad11d7}

\bibitem[{{Shi} {et~al.}(2020){Shi}, {Toshikawa}, {Cai}, {Lee}, \&
  {Fang}}]{Shi20}
{Shi}, K., {Toshikawa}, J., {Cai}, Z., {Lee}, K.-S., \& {Fang}, T. 2020, \apj,
  899, 79, \dodoi{10.3847/1538-4357/aba626}

\bibitem[{{Shi} {et~al.}(2021){Shi}, {Toshikawa}, {Lee}, {Wang}, {Cai}, \&
  {Fang}}]{Shi21}
{Shi}, K., {Toshikawa}, J., {Lee}, K.-S., {et~al.} 2021, \apj, 911, 46,
  \dodoi{10.3847/1538-4357/abe62e}

\bibitem[{{Shi} {et~al.}(2019{\natexlab{a}}){Shi}, {Lee}, {Dey}, {Huang},
  {Malavasi}, {Hung}, {Inami}, {Ashby}, {Duncan}, {Xue}, {Reddy}, {Hong},
  {Jannuzi}, {Cooper}, {Gonzalez}, {R{\"o}ttgering}, {Best}, \&
  {Tasse}}]{Shi191}
{Shi}, K., {Lee}, K.-S., {Dey}, A., {et~al.} 2019{\natexlab{a}}, \apj, 871, 83,
  \dodoi{10.3847/1538-4357/aaf85d}

\bibitem[{{Shi} {et~al.}(2019{\natexlab{b}}){Shi}, {Huang}, {Lee}, {Toshikawa},
  {Bowen}, {Malavasi}, {Lemaux}, {Cucciati}, {Le Fevre}, \& {Dey}}]{Shi192}
{Shi}, K., {Huang}, Y., {Lee}, K.-S., {et~al.} 2019{\natexlab{b}}, \apj, 879,
  9, \dodoi{10.3847/1538-4357/ab2118}

\bibitem[{{Shimakawa} {et~al.}(2018){Shimakawa}, {Kodama}, {Hayashi},
  {Prochaska}, {Tanaka}, {Cai}, {Suzuki}, {Tadaki}, \& {Koyama}}]{Shimakawa18}
{Shimakawa}, R., {Kodama}, T., {Hayashi}, M., {et~al.} 2018, \mnras, 473, 1977,
  \dodoi{10.1093/mnras/stx2494}

\bibitem[{{Snyder} {et~al.}(2012){Snyder}, {Brodwin}, {Mancone}, {Zeimann},
  {Stanford}, {Gonzalez}, {Stern}, {Eisenhardt}, {Brown}, {Dey}, {Jannuzi}, \&
  {Perlmutter}}]{Snyder12}
{Snyder}, G.~F., {Brodwin}, M., {Mancone}, C.~M., {et~al.} 2012, \apj, 756,
  114, \dodoi{10.1088/0004-637X/756/2/114}

\bibitem[{{Speagle} {et~al.}(2014){Speagle}, {Steinhardt}, {Capak}, \&
  {Silverman}}]{Speagle14}
{Speagle}, J.~S., {Steinhardt}, C.~L., {Capak}, P.~L., \& {Silverman}, J.~D.
  2014, \apjs, 214, 15, \dodoi{10.1088/0067-0049/214/2/15}

\bibitem[{{Springel} {et~al.}(2005){Springel}, {White}, {Jenkins}, {Frenk},
  {Yoshida}, {Gao}, {Navarro}, {Thacker}, {Croton}, {Helly}, {Peacock}, {Cole},
  {Thomas}, {Couchman}, {Evrard}, {Colberg}, \& {Pearce}}]{Springel05}
{Springel}, V., {White}, S. D.~M., {Jenkins}, A., {et~al.} 2005, \nat, 435,
  629, \dodoi{10.1038/nature03597}

\bibitem[{{Staab} {et~al.}(2024){Staab}, {Lemaux}, {Forrest}, {Shah},
  {Cucciati}, {Lubin}, {Gal}, {Hung}, {Shen}, {Giddings}, {Khusanova},
  {Zamorani}, {Bardelli}, {Cassara}, {Cassata}, {Chiang}, {Fudamoto},
  {Fukushima}, {Garilli}, {Giavalisco}, {Gruppioni}, {Guaita}, {Gururajan},
  {Hathi}, {Kashino}, {Scoville}, {Talia}, {Vergani}, \& {Zucca}}]{Staab24}
{Staab}, P., {Lemaux}, B.~C., {Forrest}, B., {et~al.} 2024, \mnras, 528, 6934,
  \dodoi{10.1093/mnras/stae301}

\bibitem[{{Stanford} {et~al.}(1998){Stanford}, {Eisenhardt}, \&
  {Dickinson}}]{Stanford98}
{Stanford}, S.~A., {Eisenhardt}, P.~R., \& {Dickinson}, M. 1998, \apj, 492,
  461, \dodoi{10.1086/305050}

\bibitem[{{Steidel} {et~al.}(1998){Steidel}, {Adelberger}, {Dickinson},
  {Giavalisco}, {Pettini}, \& {Kellogg}}]{Steidel98}
{Steidel}, C.~C., {Adelberger}, K.~L., {Dickinson}, M., {et~al.} 1998, \apj,
  492, 428, \dodoi{10.1086/305073}

\bibitem[{{Steidel} {et~al.}(2005){Steidel}, {Adelberger}, {Shapley}, {Erb},
  {Reddy}, \& {Pettini}}]{Steidel05}
{Steidel}, C.~C., {Adelberger}, K.~L., {Shapley}, A.~E., {et~al.} 2005, \apj,
  626, 44, \dodoi{10.1086/429989}

\bibitem[{{Suzuki} {et~al.}(2015){Suzuki}, {Kodama}, {Tadaki}, {Hayashi},
  {Koyama}, {Tanaka}, {Minowa}, {Shimakawa}, \& {Yamamoto}}]{Suzuki15}
{Suzuki}, T.~L., {Kodama}, T., {Tadaki}, K.-i., {et~al.} 2015, \apj, 806, 208,
  \dodoi{10.1088/0004-637X/806/2/208}

\bibitem[{{Suzuki} {et~al.}(2016){Suzuki}, {Kodama}, {Sobral}, {Khostovan},
  {Hayashi}, {Shimakawa}, {Koyama}, {Tadaki}, {Tanaka}, {Minowa}, {Yamamoto},
  {Smail}, \& {Best}}]{Suzuki16}
{Suzuki}, T.~L., {Kodama}, T., {Sobral}, D., {et~al.} 2016, \mnras, 462, 181,
  \dodoi{10.1093/mnras/stw1655}

\bibitem[{{Taamoli} {et~al.}(2024){Taamoli}, {Mobasher}, {Chartab}, {Darvish},
  {Weaver}, {Hemmati}, {Casey}, {Sattari}, {Brammer}, {Capak}, {Ilbert},
  {Kartaltepe}, {McCracken}, {Moneti}, {Sanders}, {Scoville}, {Steinhardt}, \&
  {Toft}}]{Taamoli24}
{Taamoli}, S., {Mobasher}, B., {Chartab}, N., {et~al.} 2024, \apj, 966, 18,
  \dodoi{10.3847/1538-4357/ad32c5}

\bibitem[{{Tadaki} {et~al.}(2013){Tadaki}, {Kodama}, {Tanaka}, {Hayashi},
  {Koyama}, \& {Shimakawa}}]{Tadaki13}
{Tadaki}, K.-i., {Kodama}, T., {Tanaka}, I., {et~al.} 2013, \apj, 778, 114,
  \dodoi{10.1088/0004-637X/778/2/114}

\bibitem[{{Tamura} {et~al.}(2009){Tamura}, {Kohno}, {Nakanishi}, {Hatsukade},
  {Iono}, {Wilson}, {Yun}, {Takata}, {Matsuda}, {Tosaki}, {Ezawa}, {Perera},
  {Scott}, {Austermann}, {Hughes}, {Aretxaga}, {Chung}, {Oshima}, {Yamaguchi},
  {Tanaka}, \& {Kawabe}}]{Tamura09}
{Tamura}, Y., {Kohno}, K., {Nakanishi}, K., {et~al.} 2009, \nat, 459, 61,
  \dodoi{10.1038/nature07947}

\bibitem[{{Tanaka} {et~al.}(2024){Tanaka}, {Onodera}, {Shimakawa}, {Ito},
  {Kakimoto}, {Kubo}, {Morishita}, {Toft}, {Valentino}, \& {Wu}}]{Tanaka24}
{Tanaka}, M., {Onodera}, M., {Shimakawa}, R., {et~al.} 2024, \apj, 970, 59,
  \dodoi{10.3847/1538-4357/ad5316}

\bibitem[{{Tang} {et~al.}(2021){Tang}, {Stark}, {Chevallard}, {Charlot},
  {Endsley}, \& {Congiu}}]{Tang21}
{Tang}, M., {Stark}, D.~P., {Chevallard}, J., {et~al.} 2021, \mnras, 503, 4105,
  \dodoi{10.1093/mnras/stab705}

\bibitem[{{Thomas} {et~al.}(2005){Thomas}, {Maraston}, {Bender}, \& {Mendes de
  Oliveira}}]{Thomas05}
{Thomas}, D., {Maraston}, C., {Bender}, R., \& {Mendes de Oliveira}, C. 2005,
  \apj, 621, 673, \dodoi{10.1086/426932}

\bibitem[{{Toshikawa} {et~al.}(2020){Toshikawa}, {Malkan}, {Kashikawa},
  {Overzier}, {Uchiyama}, {Ota}, {Ishikawa}, \& {Ito}}]{Toshikawa20}
{Toshikawa}, J., {Malkan}, M.~A., {Kashikawa}, N., {et~al.} 2020, \apj, 888,
  89, \dodoi{10.3847/1538-4357/ab5e85}

\bibitem[{{Toshikawa} {et~al.}(2012){Toshikawa}, {Kashikawa}, {Ota},
  {Morokuma}, {Shibuya}, {Hayashi}, {Nagao}, {Jiang}, {Malkan}, {Egami},
  {Shimasaku}, {Motohara}, \& {Ishizaki}}]{Toshikawa12}
{Toshikawa}, J., {Kashikawa}, N., {Ota}, K., {et~al.} 2012, \apj, 750, 137,
  \dodoi{10.1088/0004-637X/750/2/137}

\bibitem[{{Toshikawa} {et~al.}(2016){Toshikawa}, {Kashikawa}, {Overzier},
  {Malkan}, {Furusawa}, {Ishikawa}, {Onoue}, {Ota}, {Tanaka}, {Niino}, \&
  {Uchiyama}}]{Toshikawa16}
{Toshikawa}, J., {Kashikawa}, N., {Overzier}, R., {et~al.} 2016, \apj, 826,
  114, \dodoi{10.3847/0004-637X/826/2/114}

\bibitem[{{Toshikawa} {et~al.}(2024){Toshikawa}, {Wuyts}, {Kashikawa},
  {Uchiyama}, {Bremer}, {Sawicki}, {Ono}, {Kubo}, \& {Ito}}]{Toshikawa24b}
{Toshikawa}, J., {Wuyts}, S., {Kashikawa}, N., {et~al.} 2024, arXiv e-prints,
  arXiv:2404.15910, \dodoi{10.48550/arXiv.2404.15910}

\bibitem[{{Tran} {et~al.}(2020){Tran}, {Forrest}, {Alcorn}, {Yuan},
  {Nanayakkara}, {Cohn}, {Cowley}, {Glazebrook}, {Gupta}, {Kacprzak}, {Kewley},
  {Labb{\'e}}, {Papovich}, {Spitler}, {Straatman}, \& {Tomczak}}]{Tran20}
{Tran}, K.-V.~H., {Forrest}, B., {Alcorn}, L.~Y., {et~al.} 2020, \apj, 898, 45,
  \dodoi{10.3847/1538-4357/ab8cba}

\bibitem[{{Uchiyama} {et~al.}(2018){Uchiyama}, {Toshikawa}, {Kashikawa},
  {Overzier}, {Chiang}, {Marinello}, {Tanaka}, {Niino}, {Ishikawa}, {Onoue},
  {Ichikawa}, {Akiyama}, {Coupon}, {Harikane}, {Imanishi}, {Kodama},
  {Komiyama}, {Lee}, {Lin}, {Miyazaki}, {Nagao}, {Nishizawa}, {Ono}, {Ouchi},
  \& {Wang}}]{Uchiyama18}
{Uchiyama}, H., {Toshikawa}, J., {Kashikawa}, N., {et~al.} 2018, \pasj, 70,
  S32, \dodoi{10.1093/pasj/psx112}

\bibitem[{{Venemans} {et~al.}(2007){Venemans}, {R{\"o}ttgering}, {Miley}, {van
  Breugel}, {de Breuck}, {Kurk}, {Pentericci}, {Stanford}, {Overzier}, {Croft},
  \& {Ford}}]{Venemans07}
{Venemans}, B.~P., {R{\"o}ttgering}, H.~J.~A., {Miley}, G.~K., {et~al.} 2007,
  \aap, 461, 823, \dodoi{10.1051/0004-6361:20053941}

\bibitem[{{Wang} {et~al.}(2016){Wang}, {Elbaz}, {Daddi}, {Finoguenov}, {Liu},
  {Schreiber}, {Mart{\'\i}n}, {Strazzullo}, {Valentino}, {van der Burg},
  {Zanella}, {Ciesla}, {Gobat}, {Le Brun}, {Pannella}, {Sargent}, {Shu}, {Tan},
  {Cappelluti}, \& {Li}}]{Wang16}
{Wang}, T., {Elbaz}, D., {Daddi}, E., {et~al.} 2016, \apj, 828, 56,
  \dodoi{10.3847/0004-637X/828/1/56}

\bibitem[{{Wang} {et~al.}(2010){Wang}, {Cowie}, {Barger}, {Keenan}, \&
  {Ting}}]{Wang10}
{Wang}, W.-H., {Cowie}, L.~L., {Barger}, A.~J., {Keenan}, R.~C., \& {Ting},
  H.-C. 2010, \apjs, 187, 251, \dodoi{10.1088/0067-0049/187/1/251}

\bibitem[{{Weaver} {et~al.}(2023){Weaver}, {Davidzon}, {Toft}, {Ilbert},
  {McCracken}, {Gould}, {Jespersen}, {Steinhardt}, {Lagos}, {Capak}, {Casey},
  {Chartab}, {Faisst}, {Hayward}, {Kartaltepe}, {Kauffmann}, {Koekemoer},
  {Kokorev}, {Laigle}, {Liu}, {Long}, {Magdis}, {McPartland}, {Milvang-Jensen},
  {Mobasher}, {Moneti}, {Peng}, {Sanders}, {Shuntov}, {Sneppen}, {Valentino},
  {Zalesky}, \& {Zamorani}}]{Weaver23}
{Weaver}, J.~R., {Davidzon}, I., {Toft}, S., {et~al.} 2023, \aap, 677, A184,
  \dodoi{10.1051/0004-6361/202245581}

\bibitem[{{Wechsler} {et~al.}(2006){Wechsler}, {Zentner}, {Bullock},
  {Kravtsov}, \& {Allgood}}]{Wechsler06}
{Wechsler}, R.~H., {Zentner}, A.~R., {Bullock}, J.~S., {Kravtsov}, A.~V., \&
  {Allgood}, B. 2006, \apj, 652, 71, \dodoi{10.1086/507120}

\bibitem[{{Wen} {et~al.}(2022){Wen}, {An}, {Zheng}, {Shi}, {Qin}, {Gonzalez},
  {Bian}, {Xu}, {Pan}, {Tan}, {Liu}, {Fang}, {Ren}, {Zhang}, {Qiao}, \&
  {Liu}}]{Wen22}
{Wen}, R., {An}, F., {Zheng}, X.~Z., {et~al.} 2022, \apj, 933, 50,
  \dodoi{10.3847/1538-4357/ac7392}

\bibitem[{{Williams} {et~al.}(2009){Williams}, {Quadri}, {Franx}, {van Dokkum},
  \& {Labb{\'e}}}]{Williams09}
{Williams}, R.~J., {Quadri}, R.~F., {Franx}, M., {van Dokkum}, P., \&
  {Labb{\'e}}, I. 2009, \apj, 691, 1879, \dodoi{10.1088/0004-637X/691/2/1879}

\bibitem[{{Wylezalek} {et~al.}(2013){Wylezalek}, {Galametz}, {Stern}, {Vernet},
  {De Breuck}, {Seymour}, {Brodwin}, {Eisenhardt}, {Gonzalez}, {Hatch},
  {Jarvis}, {Rettura}, {Stanford}, \& {Stevens}}]{Wylezalek13}
{Wylezalek}, D., {Galametz}, A., {Stern}, D., {et~al.} 2013, \apj, 769, 79,
  \dodoi{10.1088/0004-637X/769/1/79}

\bibitem[{{Yonekura} {et~al.}(2022){Yonekura}, {Kajisawa}, {Hamaguchi},
  {Mawatari}, \& {Yamada}}]{Yonekura22}
{Yonekura}, N., {Kajisawa}, M., {Hamaguchi}, E., {Mawatari}, K., \& {Yamada},
  T. 2022, \apj, 930, 102, \dodoi{10.3847/1538-4357/ac6257}

\bibitem[{{Zavala} {et~al.}(2019){Zavala}, {Casey}, {Scoville}, {Champagne},
  {Chiang}, {Dannerbauer}, {Drew}, {Fu}, {Spilker}, {Spitler}, {Tran},
  {Treister}, \& {Toft}}]{Zavala19}
{Zavala}, J.~A., {Casey}, C.~M., {Scoville}, N., {et~al.} 2019, \apj, 887, 183,
  \dodoi{10.3847/1538-4357/ab5302}

\bibitem[{{Zentner} {et~al.}(2014){Zentner}, {Hearin}, \& {van den
  Bosch}}]{Zentner14}
{Zentner}, A.~R., {Hearin}, A.~P., \& {van den Bosch}, F.~C. 2014, \mnras, 443,
  3044, \dodoi{10.1093/mnras/stu1383}

\bibitem[{{Zhai} {et~al.}(2021){Zhai}, {Wang}, {Benson}, {Chuang}, \&
  {Yepes}}]{Zhai21}
{Zhai}, Z., {Wang}, Y., {Benson}, A., {Chuang}, C.-H., \& {Yepes}, G. 2021,
  \mnras, 505, 2784, \dodoi{10.1093/mnras/stab1539}

\bibitem[{{Zhang} {et~al.}(2022){Zhang}, {Zheng}, {Shi}, {Gao}, {Dannerbauer},
  {An}, {Shu}, {Gao}, {Wang}, {Wang}, {Cai}, {Fan}, {Fang}, {Pan}, {Liu},
  {Tan}, {Qin}, {Ren}, {Qiao}, {Wen}, \& {Liu}}]{Zhang22}
{Zhang}, Y., {Zheng}, X.~Z., {Shi}, D.~D., {et~al.} 2022, \mnras, 512, 4893,
  \dodoi{10.1093/mnras/stac824}

\end{thebibliography}
\bibliographystyle{aasjournal}



\end{document}